\documentclass[journal]{IEEEtran}
\usepackage{ifpdf}

\usepackage{cite}

\ifCLASSINFOpdf
  \usepackage[pdftex]{graphicx}
  \graphicspath{{figs/}{}}
  \DeclareGraphicsExtensions{.pdf,.jpeg,.png}
  \usepackage{epstopdf}
\else
  
  \usepackage[dvips]{graphicx}
  \graphicspath{{figs/}}
  \DeclareGraphicsExtensions{.eps}
\fi
\usepackage[cmex10]{amsmath}
\usepackage{amssymb}

\usepackage{array}

\usepackage[normalem]{ulem}	

\usepackage{verbatim}

\usepackage{color}

\usepackage{version}
\usepackage{siunitx}

\excludeversion{notes}

\begin{document}
%
\title{The Complex Permeability of Split-Ring Resonator Arrays Measured at Microwave Frequencies}
%
%
%


\author{Sabrina~L.~Madsen 
       and~Jake~S.~Bobowski
\thanks{Manuscript submitted on \today.}
\thanks{S.L.~Madsen and J.S.~Bobowski are with the Department
of Physics, University of British Columbia, Kelowna,
BC, V1V 1V7 Canada (e-mail: \mbox{jake.bobowski@ubc.ca}).}
}

%
%


\markboth{Madsen and Bobowski: The complex permeability of split-ring resonator arrays measured at microwave frequencies}{}
%



\maketitle

\begin{abstract}
We have measured the relative permeability of split-ring resonator (SRR) arrays used in metamaterials designed to have $\boldsymbol{\mu^\prime< 0}$ over a narrow range of microwave frequencies.  The SRR arrays were loaded into the bore of a loop-gap resonator (LGR) and reflection coefficient measurements were used to determine both the real and imaginary parts of the array's effective permeability.  Data were collected as a function of array size and SRR spacing.  The results were compared to those obtained from continuous extended split-ring resonators (ESRRs).  The arrays of planar SRRs exhibited enhanced damping and a narrower range of frequencies with $\boldsymbol{\mu^\prime<0}$ when compared to the ESRRs.  The observed differences in damping, however, were diminished considerably when the array size was expanded from a one-dimensional array of $\boldsymbol{N}$ SRRs to a $\boldsymbol{2\times 2\times N}$ array. Our method can also be used to experimentally determine the effective permeability of other metamaterial designs.
\end{abstract}


%
\IEEEpeerreviewmaketitle

\section{\label{sec:intro}Introduction}
%
%
%
%

\IEEEPARstart{T}{his} paper describes experimental measurements of the relative permeability $\mu^\prime-j\mu^{\prime\prime}$ of arrays of planar split-ring resonators (SRRs) at microwave frequencies.  The SRR array was first proposed by Pendry {\it et al}.\ as an artificial structure that could be engineered to have magnetic properties ($\mu^\prime<0$) not found in natural materials~\cite{Pendry:1999}.  A short time later, a metamaterial with a negative index of refraction was fabricated from a lattice of SRRs and conducting rods.  Negative index materials (NIMs) have an effective permeability $\mu$ and permittivity $\varepsilon$ that are simultaneously negative over a narrow band of microwave frequencies~\cite{Smith:2000, Shelby:2001}.  Metamaterials have generated a lot of interest due to their potential use in exotic applications~\cite{Padilla:2006, Alitalo:2009}.

There are two primary motivators for our present work.  First, Pendry's initial negative-permeability calculation considered a pair of concentric conducting sheets arranged in a split-ring geometry.  We will refer to this structure as an extended split-ring resonator (ESRR)~\cite{Bobowski:2018}.  Pendry {\it et al}.\ argued that arrays of planar SRRs should approximate the behavior of the ESRR and an expression for the permeability of the array was derived.  However, the expression was valid only for a certain set of geometrical constraints that are difficult to satisfy in practice.  For example, their estimate of the SRR capacitance required the inner radius of the concentric rings to be greater than the spacing between SRRs.  In practical realizations of a three-dimensional (3D) metamaterial, on the other hand, the size of a unit cell is typically greater than the SRR outer diameter~\cite{Pendry:1999}.

The second driver of our work has been the fact that extracting meaningful effective parameters from metamaterials has proven to be challenging, both numerically and experimentally.  The most widely used technique is known as the Nicolson-Ross-Weir (NRW) method which involves illuminating the metamaterial with plane waves and analyzing the resulting scattering parameters~\cite{Nicolson:1970, Weir:1974}.  Using this method, there have been many reports in which either $\varepsilon^{\prime\prime}$ or $\mu^{\prime\prime}$ have been found to be negative near the resonant frequency of the metamaterial.  A value of $\xi^{\prime\prime}<0$, where $\xi$ is either $\varepsilon$ or $\mu$, is counter to physical intuition because it represents a gain in energy.  This type of behavior has been termed an anti-resonant response and it has been observed in both numerical and experimental studies~\cite{Koschny:2003,Alitalo:2013,Liu:2016, Itoh:2018}.

Koschny {\it et al}.\ argued that $\xi^{\prime\prime}<0$ does not violate any physical principle so long as there is net dissipation of the electromagnetic (EM) field by the metamaterial~\cite{Koschny:2003}.  Furthermore, they argued that $\xi^{\prime\prime}<0$ is an intrinsic property of metamaterials caused by the spatial periodicity of the composite structures~\cite{Koschny:2005}.  Alitalo {\it et al}.\ applied the NRW method to aperiodic metamaterial structures, both experimentally and numerically, and found that values of $\xi^{\prime\prime}<0$ persisted.  They interpreted their results as pointing to a deficiency in the NRW method when applied to microwave metamaterials~\cite{Alitalo:2013}.  This view was supported by Woodley and Mojahedi who applied the NRW method to an array of dielectric spheres for which there is an analytical solution that shows that $\varepsilon^{\prime\prime}$ is positive at all frequencies.  In contrast, their numerical studies using the NRW method found $\varepsilon^{\prime\prime}<0$ over a narrow frequency band~\cite{Woodley:2010}.  

Given the questions over the reliability of the NRW method when applied to metamaterials, alternative methods for extracting the effective parameters of these structures need to be developed.  Kong and coworkers have developed a method in which the metamaterial is loaded into a rectangular waveguide operating in the TE$_{10}$ mode and the complex reflection and transmission coefficients are analyzed.  As with the NRW retrieval method, these authors also report an anti-resonant response~\cite{Chen:2006, Zhang:2008}.

Marqu\'es {\it et al}.~\cite{Marques:2003} have developed a method for measuring the resonant frequency and magnetic polarizability of a single resonant element used in the construction of metamaterials.  Their method places the resonator inside of a circular aperture within a rectangular waveguide.  They then make use of a Lorentz local field theory to deduce the effective parameters (permittivity and permeability) of a material made from a regular array of the resonant elements~\cite{Marques:2003}.  Our contribution is distinct from that of Marqu\'es {\it et al}.\ because our objective was to develop and implement an entirely new experimental method capable of directly extracting the complex permeability of complete metamaterial structures consisting of numerous resonant elements.

A unique feature of our method is that it exposes the material under test to an RF magnetic field while keeping its exposure to RF electric fields to a minimum.  As pointed out by Koschny {\it et al}., $\mu^{\prime\prime}>0$ is required in this situation to ensure net energy dissipation~\cite{Koschny:2003}.  Therefore, a material found to have $\mu^{\prime\prime}<0$ when exposed to EM plane waves would have to have the unusual property that its magnetic response changes when exposed to a uniform RF magnetic field.  To the best of our knowledge, such a property has not been experimentally demonstrated in any of the existing microwave metamaterials.

In this paper we describe the use of loop-gap resonators (LGRs) to experimentally determine the permeability of  both SRR arrays of various sizes and assemblies of ESRRs.  The paper is organized as follows: Section~\ref{sec:methods} provides an overview of the experimental methods.  In particular, it describes how reflection coefficient measurements of an SRR-loaded LGR are used to determine the permeability of the SRR array. In Section~\ref{sec:design}, we present the design of the LGRs, SRRs, and ESSRs used in our experiments. Section~\ref{sec:data} shows reflection coefficient measurements and the corresponding extracted permeabilities of SRR arrays and ESRRs.  In Section~\ref{sec:discussion} we discuss the results and Section~\ref{sec:conclusions} summarizes the main conclusions.

\section{\label{sec:methods}Experimental Methods}
Recently, we have used LGRs to measure electric~\cite{Bobowski:2013, Bobowski:2017} and magnetic~\cite{Dubreuil:2019} material properties at microwave frequencies.  LGRs are electrically-small resonators that can be modeled, approximately, as lumped-element $LRC$ circuits~\cite{Hardy:1981, Froncisz:1982}.  In their simplest form, LGRs consist of a hollow conducting tube with a narrow slit cut along its length.  The slit dimensions set the effective capacitance of the resonator and the bore dimensions determine its inductance.  Dissipation is determined by the resistivity of the conductor and energy losses via EM radiation~\cite{Bobowski:2013}.  Typically, a coupling loop, made by short circuiting the center conductor of a coaxial cable to its outer conductor, is used to couple signals into and out of the LGR.

As described in~\cite{Bobowski:2018}, the reflection coefficient $\left\vert S_{11}\right\vert$ of an inductively-coupled LGR is determined by the inductance $L_1$ of the coupling loop, the LGR resonant frequency $f_0$ and quality factor $Q_0$, and $M_0^2/R_0$, where $M_0$ is the mutual inductance between the coupling loop and the LGR bore, and $R_0$ is the effective resistance of the LGR.  The value of $L_1$ can be determined from $S_{11}$ measurements when the coupling loop is isolated from the LGR (i.e.\ the $M_0\to 0$ limit).  Once $L_1$ is known, fits to the frequency dependence of $\left\vert S_{11}\right\vert$ of the coupled LGR can be used to determine $f_0$, $Q_0$, and $M_0^2/R_0$.

When the bore of the LGR is partially filled with a magnetic material, the inductance and, therefore, resonance of the LGR is modified by the complex permeability \mbox{$\mu^\prime-j\mu^{\prime\prime}$} of the material under test.  If $\mu^\prime$ and $\mu^{\prime\prime}$ are approximately constant over the frequency range of interest, then $\mu^\prime$ is largely determined from the shift in resonant frequency and $\mu^{\prime\prime}$ by the change in quality factor~\cite{Bobowski:2015, Dubreuil:2019}.
 
In the case of a metamaterial, however, $\mu^\prime$ and $\mu^{\prime\prime}$ are expected to be strong functions of frequency near $f_0$ of the LGR.  Therefore, to determine the permeability from a fit to an $\left\vert S_{11}\right\vert$ measurement, a model for the frequency dependence of \mbox{$\mu_\mathrm{eff}=\mu^\prime-j\mu^{\prime\prime}$} must be adopted.  For our measurements of SRR arrays, we use the model Pendry {\it et al}.\ originally calculated for ESRRs
\begin{equation}
\mu_\mathrm{eff}=1-\frac{1-\left(f_\mathrm{s}/f_\mathrm{p}\right)^2}{1-\left(f_\mathrm{s}/f\right)^2-j\left[\gamma/\left(2\pi f\right)\right]}\label{eq:1}
\end{equation} 
where $f_\mathrm{s}$ is the ESRR resonant frequency, $f_\mathrm{p}$ is called the magnetic plasma frequency, and $\gamma$ is a damping constant~\cite{Pendry:1999}.  Note that, provided $f_\mathrm{p}>f_\mathrm{s}$, this model guarantees that \mbox{$\mu^{\prime\prime}>0$} at all frequencies.  Furthermore, it satisfies the Kramers-Kronig relations for the real and imaginary components of the magnetic susceptibility, \mbox{$\chi_\mathrm{eff}=\mu_\mathrm{eff}-1$}~\cite{Sharnoff:1964, Szabo:2010}.

In the measurements presented in this paper, we first used the $\left\vert S_{11}\right\vert$ frequency dependence of an LGR with an empty bore to determine $f_0$, $Q_0$, and $M_0^2/R_0$.  Next, the bore of the LGR was partially loaded with an SRR array.  Assuming a known filling factor $\eta$, a second $\left\vert S_{11}\right\vert$ measurement was made and a fit to its frequency dependence was used to extract the parameters $f_\mathrm{s}$, $f_\mathrm{p}$, and $\gamma$ that characterize the permeability of the SRR array.  All reflection coefficient measurements were made using an Agilent E5061A vector network analyzer (VNA).

In \cite{Bobowski:2018}, the expected frequency dependence of $\left\vert S_{11}\right\vert$ for an LGR whose bore is partially filled with a negative-permeability material was calculated.  A $\mu_\mathrm{eff}$ given by (\ref{eq:1}) was assumed for the material filling the bore and the calculation revealed a reflection coefficient with a double resonance.  The frequency splitting of the pair of resonances as well as their relative amplitudes and widths were shown to depend on the filling factor $\eta$ and the SRR permeability parameters $f_\mathrm{s}$, $f_\mathrm{p}$, and $\gamma$.  We emphasize that the resonant frequency of the metamaterial permeability does not correspond to the either of the resonances in the $\left\vert S_{11}\right\vert$ frequency response.  The $\left\vert S_{11}\right\vert$ nulls are set by the LGR resonant frequency $f_0$, the filling factor, and the effective permeability of the metamaterial.  Finally,~\cite{Bobowski:2018} also presented a proof-of-principle sequence of measurements using a crude ESRR and an existing toroidal LGR \cite{Bobowski:2016} to demonstrate the validity of the experimental method.

In this paper, we apply the experimental methods developed in \cite{Bobowski:2018} to arrays of SRRs of varying sizes using LGRs specifically designed for the measurements.  Our first objective was to verify that the model in (\ref{eq:1}) could be used to successfully interpret our SRR array measurements and then to use the results to characterize the SRR permeability as a function of array size.  Secondly, we wanted to experimentally test to what extent SRR arrays approximate the behavior of ESRRs, which was one of the key insights originally proposed by Pendry {\it et al}.~\cite{Pendry:1999}.

\begin{figure}
\begin{tabular}{c}
(a)\quad\includegraphics[height=4cm]{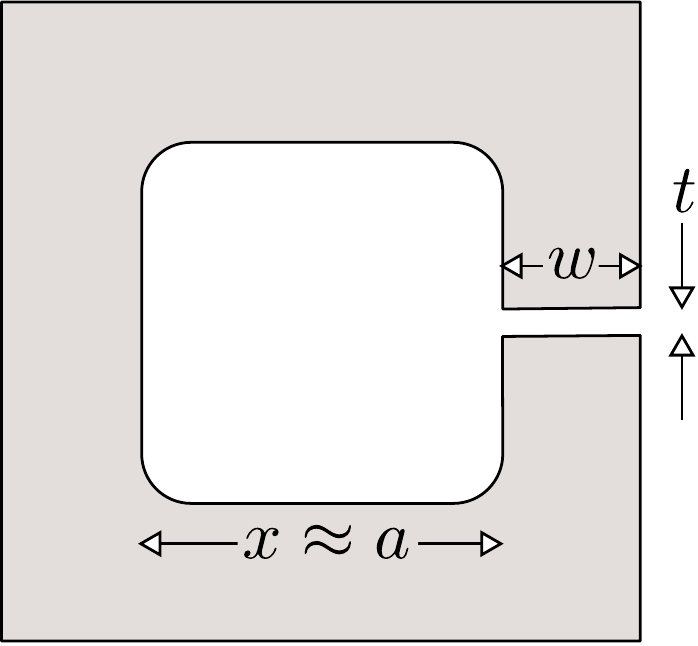}\\
~\\
(b)\quad\includegraphics[height=4cm]{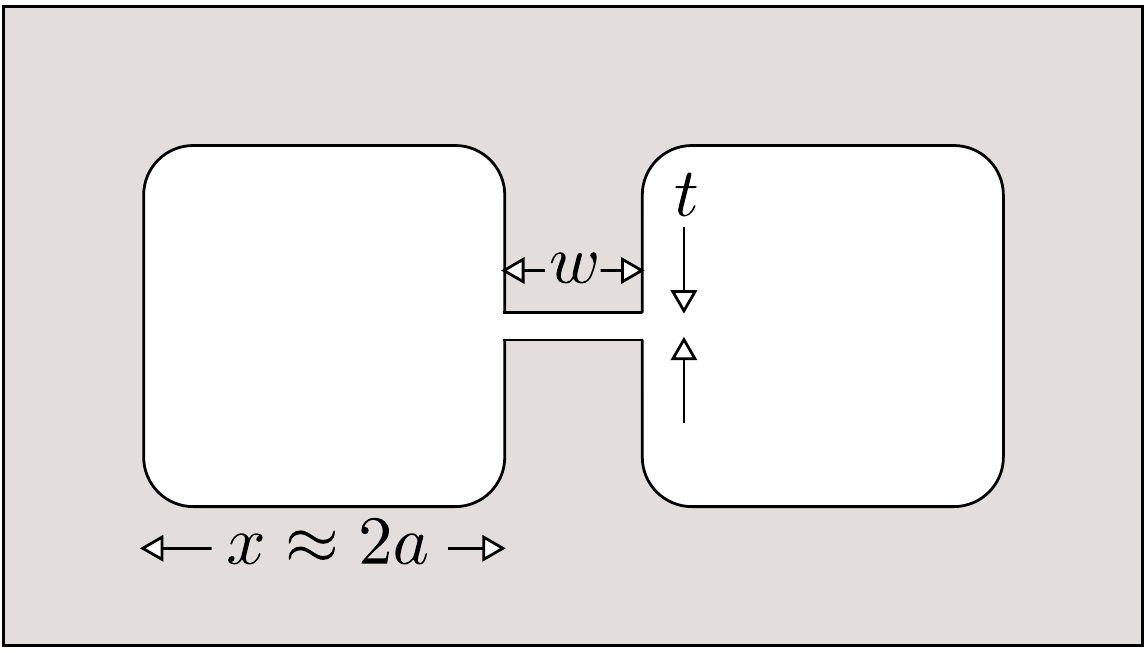}\\
~\\
(c)\quad\includegraphics[width=0.8\columnwidth]{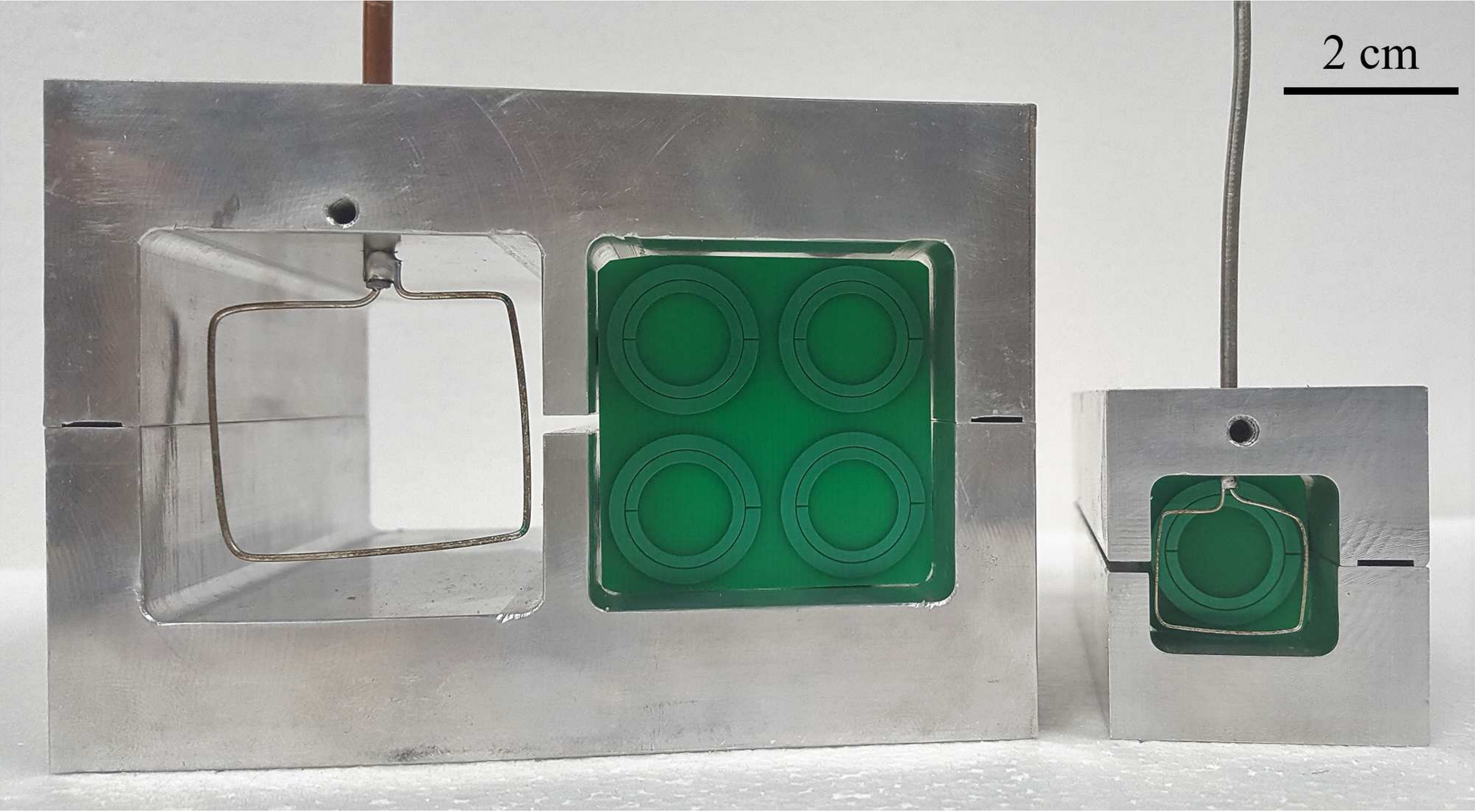}\\
~\\
(d)\quad\includegraphics[width=0.8\columnwidth]{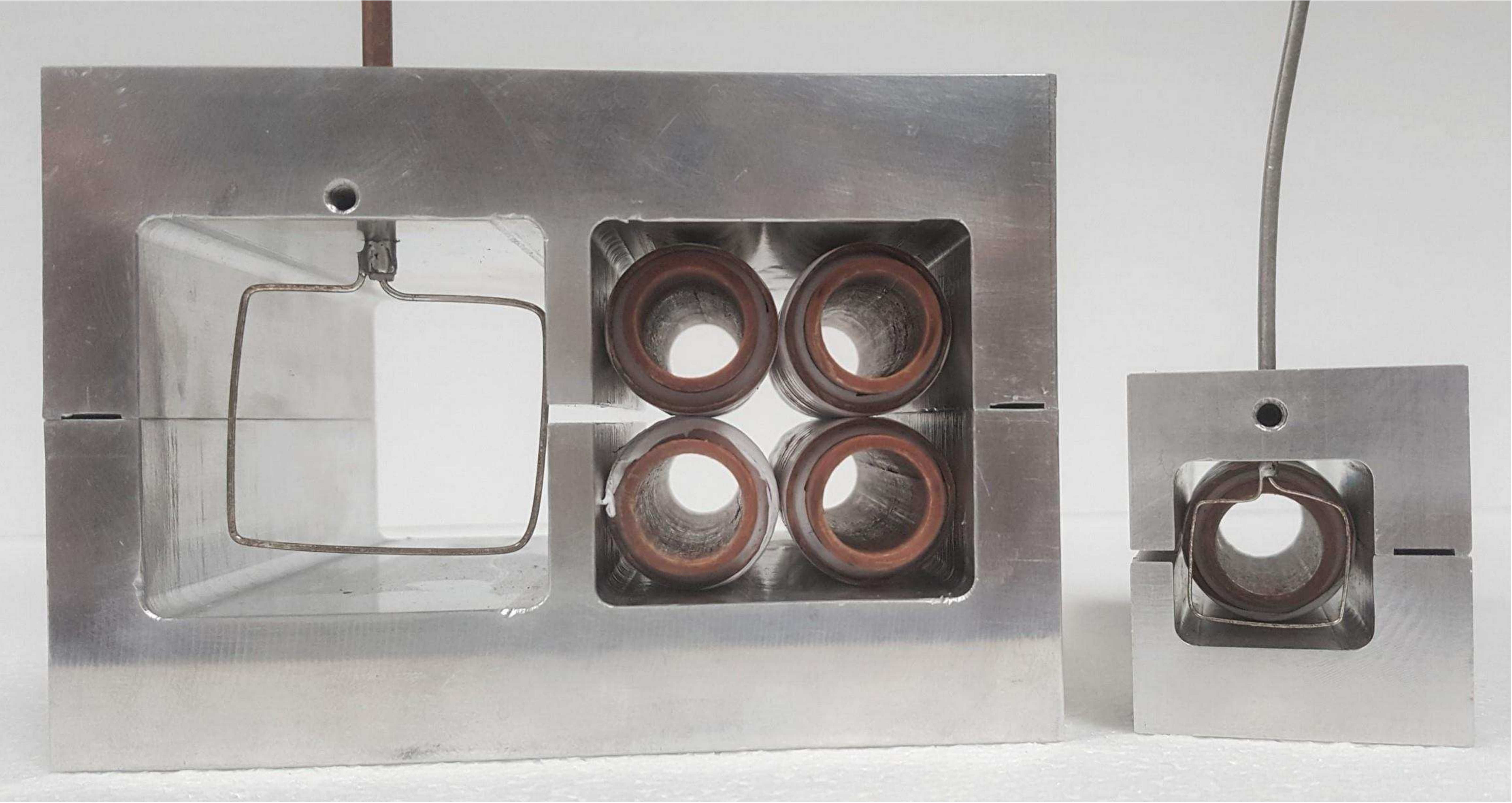}
\end{tabular}
\caption{\label{fig:Fig1}(a) Schematic drawing of the 1-loop-1-gap LGR cross section with the critical dimensions labeled (not to scale).  The resonator length was \SI{112}{\milli\meter}, $x=\SI{21.6}{\milli\meter}$, $w=\SI{5.0}{\milli\meter}$, and $t=\SI{1.3}{\milli\meter}$. (b) Schematic drawing of the 2-loop-1-gap LGR cross section (not to scale). The resonator length was \SI{112}{\milli\meter}, $x=\SI{42.7}{\milli\meter}$, $w=\SI{5.0}{\milli\meter}$, and $t=\SI{2.0}{\milli\meter}$.  (c) Photograph of the 2-loop-1-gap LGR loaded with a \mbox{$2\times 2\times N$} SRR array and the 1-loop-1-gap LGR loaded with a 1D array of $N$ SRRs.  (d) Photograph of the 2-loop-1-gap LGR loaded with a \mbox{$2\times 2$} assembly of ESRRs and the 1-loop-1-gap LGR loaded with a single ESRR.  The inductive coupling loops, made to occupy as much of the bore area as possible, are also visible in (c) and (d).
}
\end{figure}

\begin{figure*}
\begin{tabular}{cc}
{\begin{tabular}[b]{c} (a)\quad\includegraphics[height=6.5cm]{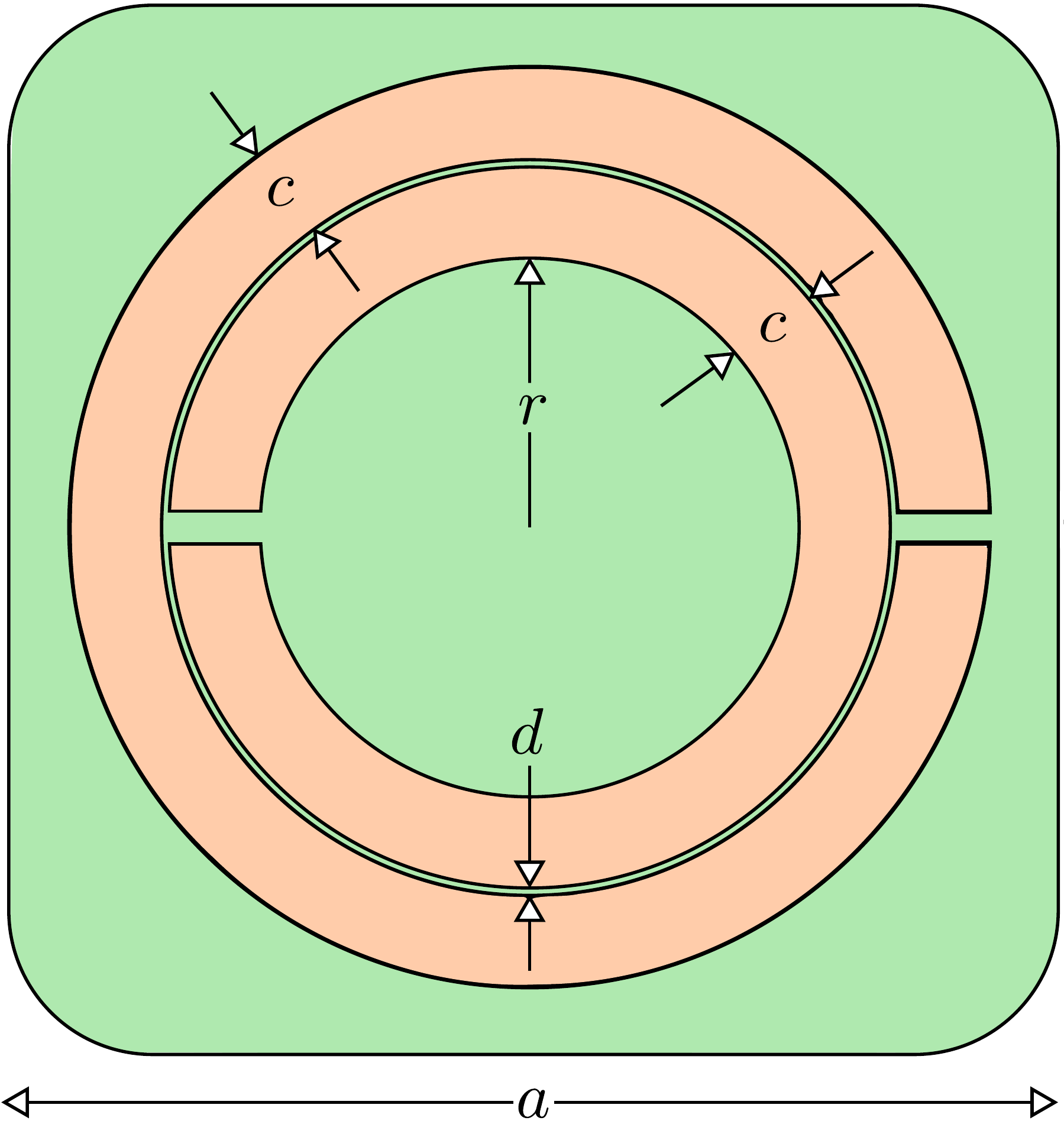}\\ ~\\~ \end{tabular}} & (b)\includegraphics[height=8cm]{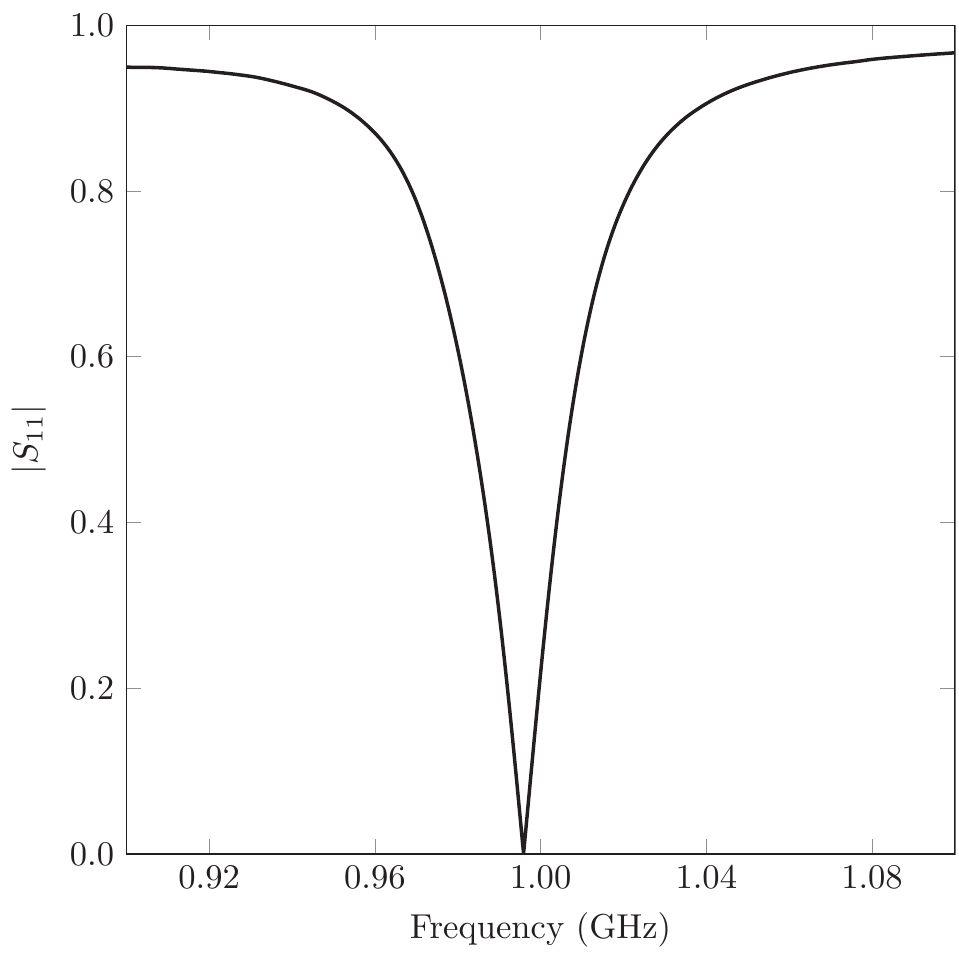}\\
~ & ~\\
\qquad {\begin{tabular}[b]{c} (c)\quad~~~\includegraphics[height=5cm]{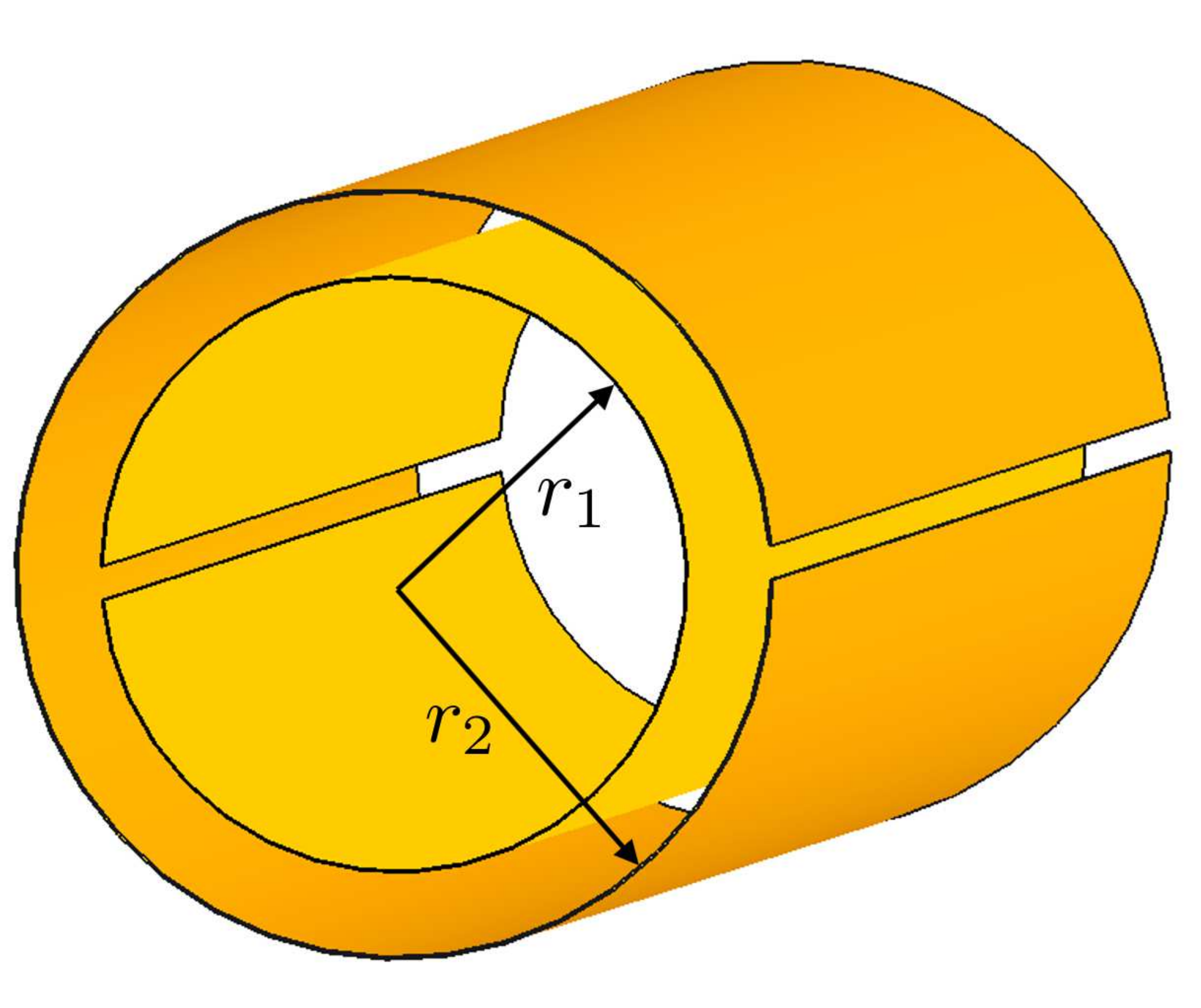}\quad~~\\ (d)~\includegraphics[width=0.8\columnwidth]{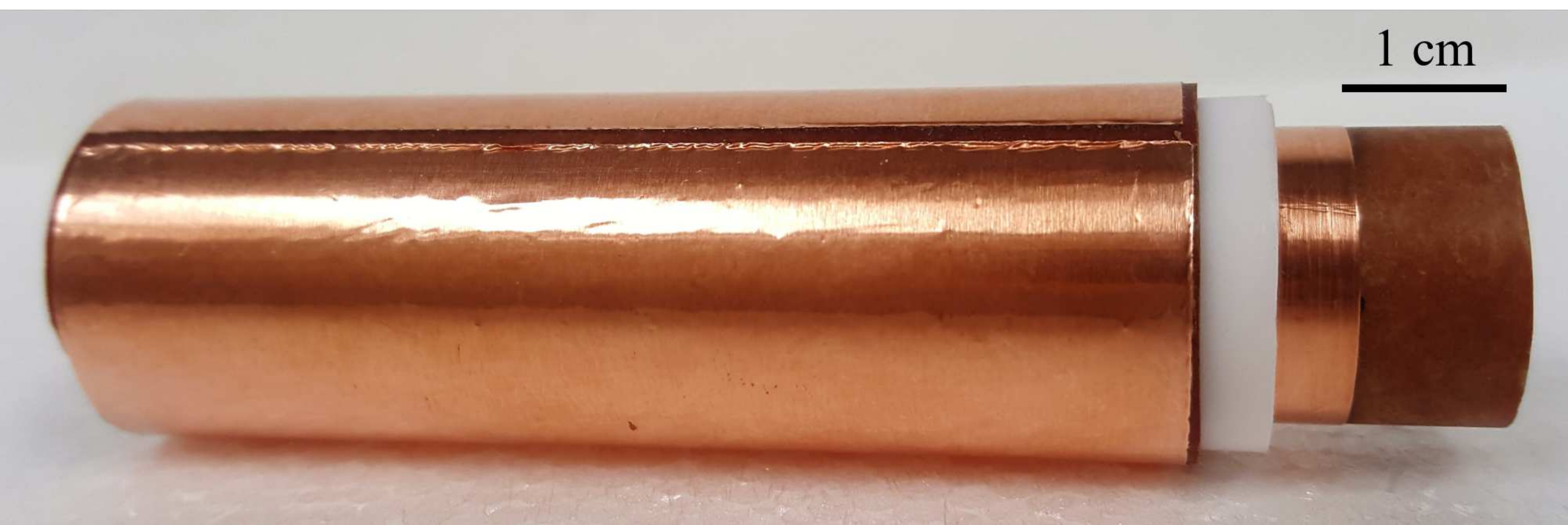}\\~\end{tabular}} & (e)\includegraphics[height=8cm]{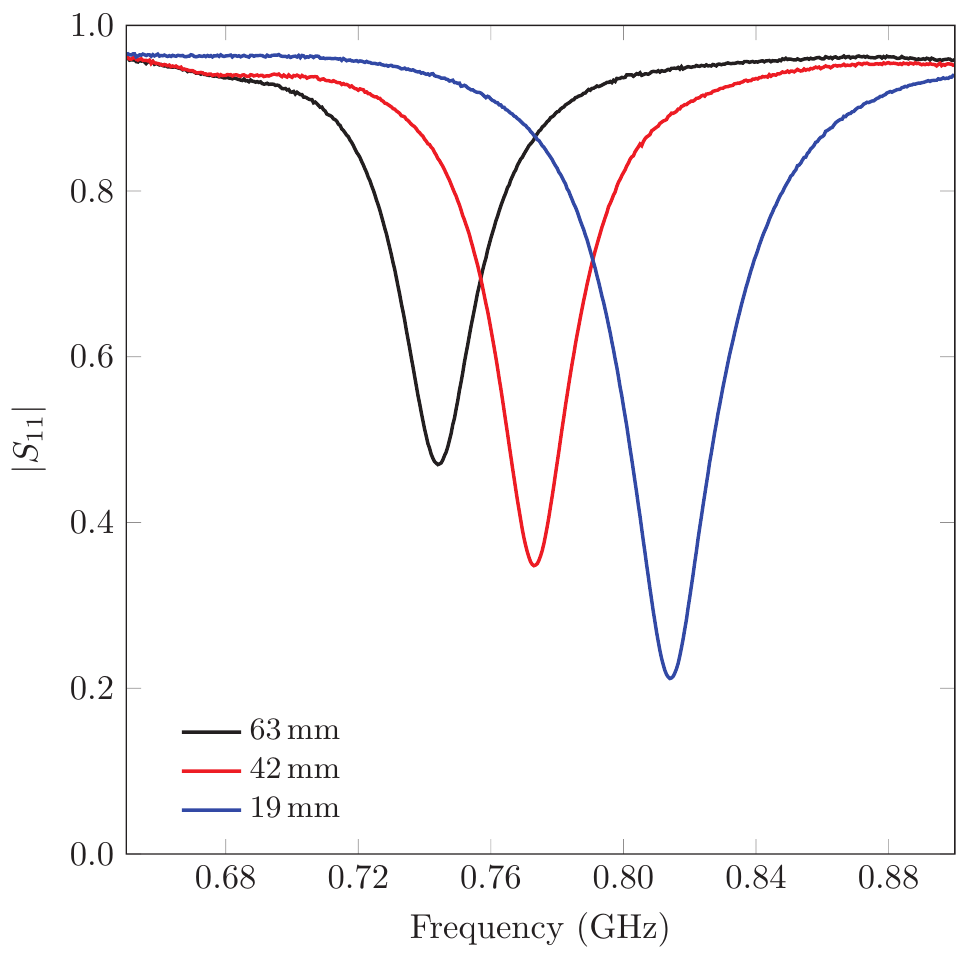}
\end{tabular}
\caption{\label{fig:Fig2}(a) Schematic drawing of the SRR geometry with the critical dimensions labeled (not to scale).  The green square with rounded corners represents the FR-4 circuit board.  The thickness of the circuit board was $t_\mathrm{SRR}=\SI{1.54}{\milli\meter}$ and $a=\SI{21.0}{\milli\meter}$, $r=\SI{5.56}{\milli\meter}$, $c=\SI{1.91}{\milli\meter}$, and $d=\SI{0.15}{\milli\meter}$.  (b)  $\left\vert S_{11}\right\vert$ as a function of frequency for the SRR.  (c) Schematic drawing of the ESRR geometry with the critical dimensions labeled (not to scale).  The ESRRs had $r_1=\SI{7.9}{\milli\meter}$, $r_2=\SI{10.3}{\milli\meter}$, and lengths $\ell_\mathrm{ESRR}$ of \SI{63}{\milli\meter}, \SI{42}{\milli\meter}, and \SI{19}{\milli\meter}. (d) Photograph of an ESRR made, in part, using copper tape.  The assembly has been pulled apart to expose the different layers of the resonator. (e) $\left\vert S_{11}\right\vert$ as a function of frequency for ESRRs of three different lengths.}
\end{figure*}

\section{\label{sec:design}LGR, SRR, \& ESRR Designs}
\subsection{\label{sec:LGR}1-Loop-1-Gap \& 2-Loop-1-Gap LGRs}
Figure~\ref{fig:Fig1} shows the designs of the inductively coupled~\cite{Rinard:1993} 1-loop-1-gap and 2-loop-1-gap~\cite{Froncisz:1986} LGRs used in our experiments.  LGRs provided good isolation between the electric and magnetic fields which are concentrated within the gap and bore of the resonator, respectively. The critical dimensions of the 1-loop-1-gap LGR resonator are given in Fig.~\ref{fig:Fig1}(a).  The corners of the square bore were rounded to avoid large current densities in these regions.  The radius of the corners was \SI{2.38}{\milli\meter} (\SI{3/32}{inch}).  Modeling the LGR as a series $LRC$ circuit predicts a single resonance with a resonant frequency given by~\cite{Bobowski:2018}
\begin{equation}
f_0\approx\frac{1}{2\pi}\frac{c}{x}\sqrt{\frac{t}{w}}\approx\SI{1}{\giga\hertz}.
\end{equation} 
Figure~\ref{fig:Fig1} also shows the design of the 2-loop-1-gap LGR and the dimensions of the resonator used in our experiments are given in Fig.~\ref{fig:Fig1}(b).  The 2-loop-1-gap LGR can be designed to have a resonant frequency that is similar to that of the 1-loop-1-gap LGR, while having a bore with a significantly larger cross-sectional area.  The 2-loop-1-gap LGR, therefore, allows us to experiment with larger arrays of SRRs.  As shown in Figs.~\ref{fig:Fig1}(c) and (d), the 1-loop-1-gap LGR was used to study one-dimensional (1D) arrays of $N$ SRRs and single ESRRs, while the 2-loop-1-gap LGR was used to study $2\times 2\times N$ arrays of SRRs and $2\times 2$ assemblies of ESRRs.  Investigating larger array sizes was critical as it allowed us to explore the importance of intraplane coupling between SRRs.    

In Figs.~\ref{fig:Fig1}(a) and (b), $a$ characterizes the size of an SRR.  The cross-sectional area of the LGR bore occupied by a single SRR is given by $a^2$.  Using the methods described in Section~\ref{sec:methods}, the empty-bore 1-loop-1-gap and 2-1oop-1-gap LGRs were characterized.  The parameter values extracted from the fits to the $\left\vert S_{11}\right\vert$ measurements (not shown) are given in Table~\ref{tab:Tab1}.  All fits in this paper were performed in MATLAB using a nonlinear fitting routine that employs the Levenberg-Marquardt algorithm.  The routine was initiated by providing starting values for the model parameters.  In the case of the empty-bore LGRs, reasonable starting values were estimated from the resonator geometry.
\begin{table}
\caption{\label{tab:Tab1}Table of the parameter values extracted from fits to $\left\vert S_{11}\right\vert$ of the 1-loop-1-gap and 2-loop-1-gap LGRs.  These results were obtained while the LGR bores were empty.}
\begin{tabular}{rccc}
~ & 1-loop-1-gap & \quad & 2-loop-1-gap\\
\\[-1em]
\hline\hline
\\[-1em]
$L_1$ (nH)\quad~ & 42.5 & ~ & 87.0\\
$M_0^2/R_0$ (nH$^2$/m$\Omega$)\quad~ & 0.053 & ~ & 0.528\\
$f_0$ (MHz)\quad~ & 858 & ~ & 652\\
$Q_0$\quad~ & 49 & ~ & 102
\end{tabular}
\end{table}

\subsection{\label{sec:SRR}SRRs \& ESSRs}
The SRRs used in our experiments were fabricated commercially on FR-4 printed circuit board which has a dielectric constant of approximately \SI{4.5}{} at \SI{1}{\giga\hertz}~\cite{Djordjevic:2001}.  The design and dimensions of the SRRs are presented in Fig.~\ref{fig:Fig2}(a).  The resonance of the SRR was determined by positioning a coupling loop above the plane of the SRR and using the VNA to measure the resulting reflection coefficient.  The strength of the coupling was easily tuned by adjusting the height of the loop above the SRR.  The data in Fig.~\ref{fig:Fig2}(b) were taken after arbitrarily positioning the coupling loop to achieve critical coupling such that $\left\vert S_{11}\right\vert=0$ at the resonant frequency~\cite{Rinard:1993}.  As shown in Fig.~\ref{fig:Fig1}(c), $2\times 2$ planes of the SRRs were fashioned from a printed circuit board of area $\left(2a\right)^2$ with one SRR in each quadrant of the square surface.

The conceptual design of the extended split-ring resonator (ESRR) is shown in Fig.~\ref{fig:Fig2}(c) and the practical realization is pictured in Fig.~\ref{fig:Fig2}(d).  The concentric conducting sheets were made using \SI[number-unit-product=\text{-}]{0.09}{\milli\meter} thick copper tape.  The inner sheet was wrapped around a paper Bakelite tube with an outer radius of $r_1=\SI{7.9}{\milli\meter}$ (\SI{5/16}{inch}).  The copper sheet was cut such that its ends did not overlap so as to avoid forming a closed conducting loop.  A second copper-tape-wrapped paper Bakelite tube with outer radius $r_2=\SI{10.3}{\milli\meter}$ (\SI{13/32}{inch}) was slid over top of the first tube.  As discussed in Section~\ref{sec:data}, when loaded into an LGR bore, the nonmagnetic paper Bakelite tubes have negligible effect on the LGR resonant frequency and quality factor.  However, the outer tube, with a wall thickness of \SI{0.9}{\milli\meter}, occupies part of the space between the conducting sheets.  As a result, its dielectric constant and loss tangent contribute to the capacitance and damping constant of the ESRR.  To ensure that the pair of tubes shared a common axis, the space between them was filled by wrapping the inner tube 3.5 times with a continuous length of \SI[number-unit-product=\text{-}]{0.38}{\milli\meter} thick Teflon sheet.  Sets of four nearly-identical ESRRs of three different lengths were made.  The resonances of these structures were characterized by inserting a coupling loop into the bore formed by the inner conducting sheet and using the VNA to measure the frequency dependence of $\left\vert S_{11}\right\vert$.  As shown in Fig.~\ref{fig:Fig2}(e), the ESRR resonant frequency decreased as the resonator length was increased.  The curvature of the magnetic field lines near the ends of the ESRRs results in a reduced magnetic flux in these regions which lowers the overall effective inductance of the resonator.  This effect is more pronounced in short ESRRs such that the resonant frequency increases with decreasing length~\cite{Hardy:1981}.

The ESRRs were wrapped in Teflon tape before they were loaded into the LGRs so as to avoid electrical shorts to the LGR bore and, in the case of the $2\times 2$ assemblies, to one another.  As shown in Fig.~\ref{fig:Fig1}(d), the $2\times 2$ structures were made by placing a row of two ESRRs in the LGR bore and then stacking a second row of ESRRs on top of the first.  The thickness of the outer layer of Teflon tape was chosen such that the ESRRs uniformly filled the cross-section of the LGR bore.

\section{\label{sec:data}Data \& Analysis}
We now present our permeability measurements of SRR arrays and ESRRs.  When SRRs were loaded into the bore of an LGR, the spacing between adjacent resonator planes was set using spacers made from paper Bakelite tubes.  The tubes had an outside diameter of \SI{20.6}{\milli\meter} (\SI{13/16}{inch}) and a wall thickness of \SI{0.9}{\milli\meter}.  Placing only a \SI[number-unit-product=\text{-}]{21}{\milli\meter} long Bakelite spacer in the bore of the 1-loop-1-gap LGR changed the resonant frequency by just \SI{0.7}{\percent} and, to within experimental uncertainty, left the quality factor unchanged.

\begin{figure*}
\begin{tabular}{cc}
(a)\includegraphics[height=7.5cm]{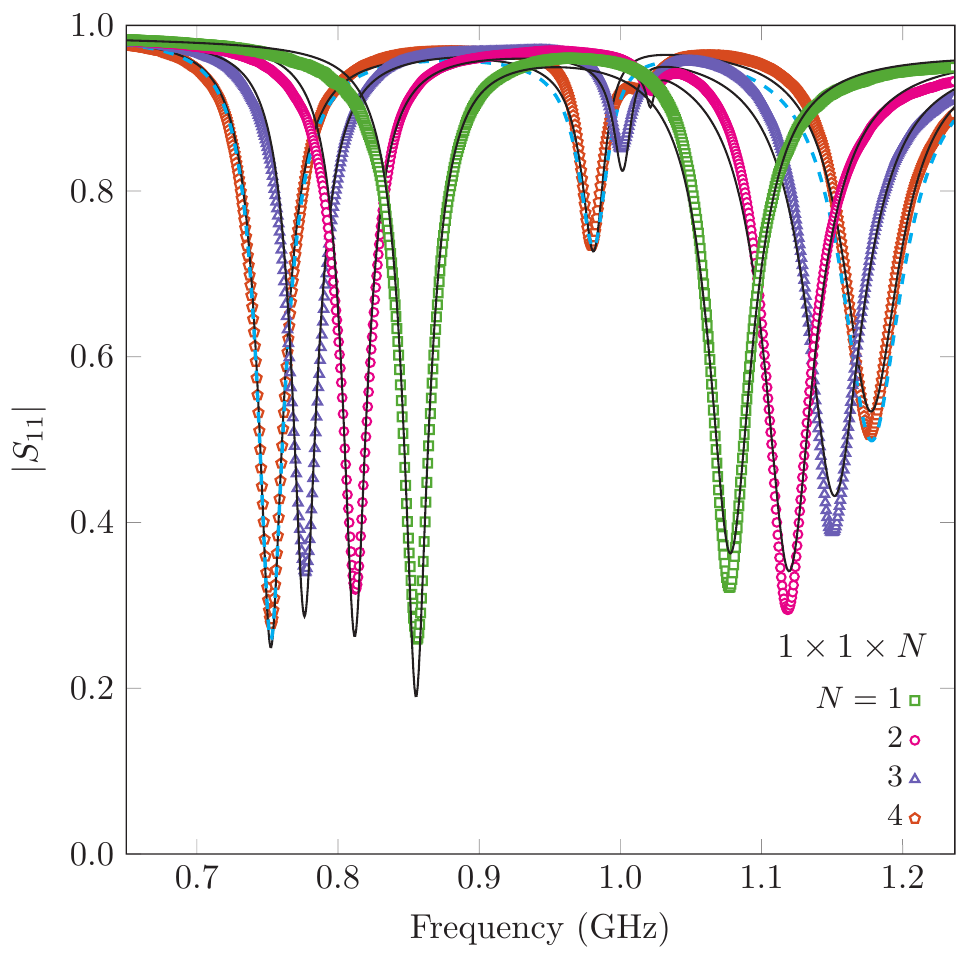} & \qquad (b)\includegraphics[height=7.5cm]{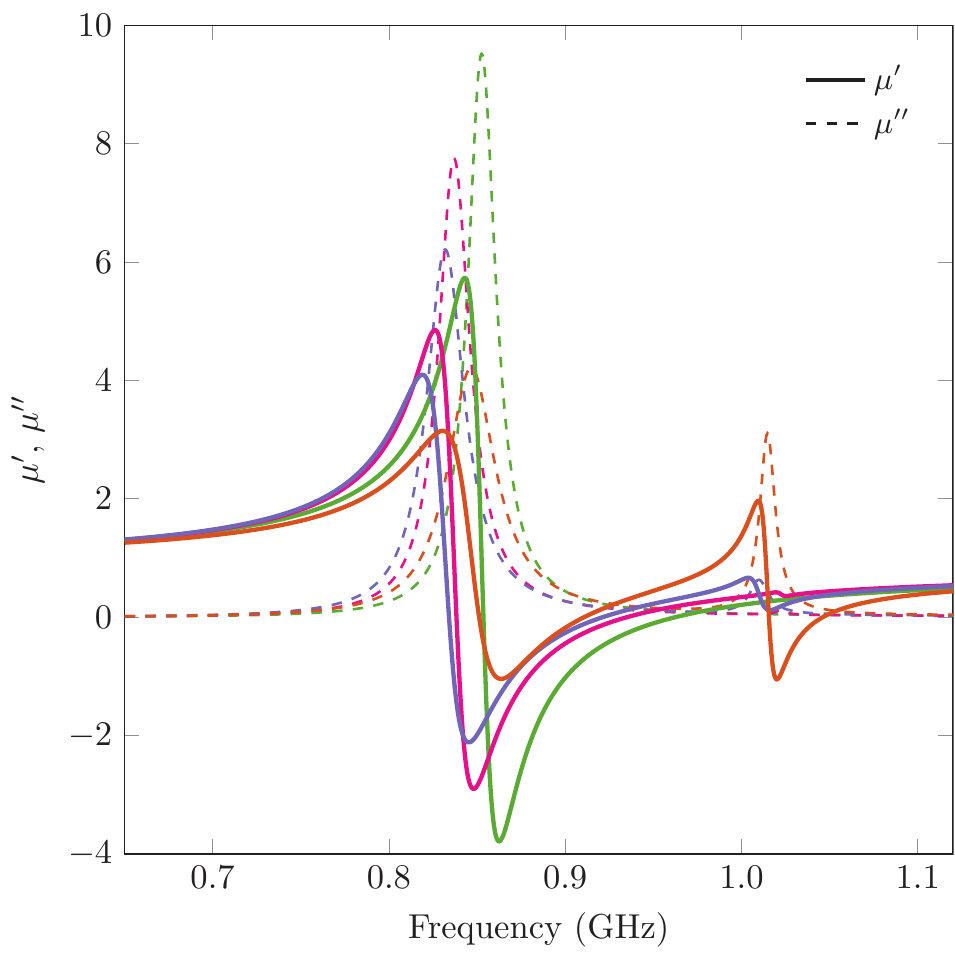}\\
(c)\includegraphics[height=7.5cm]{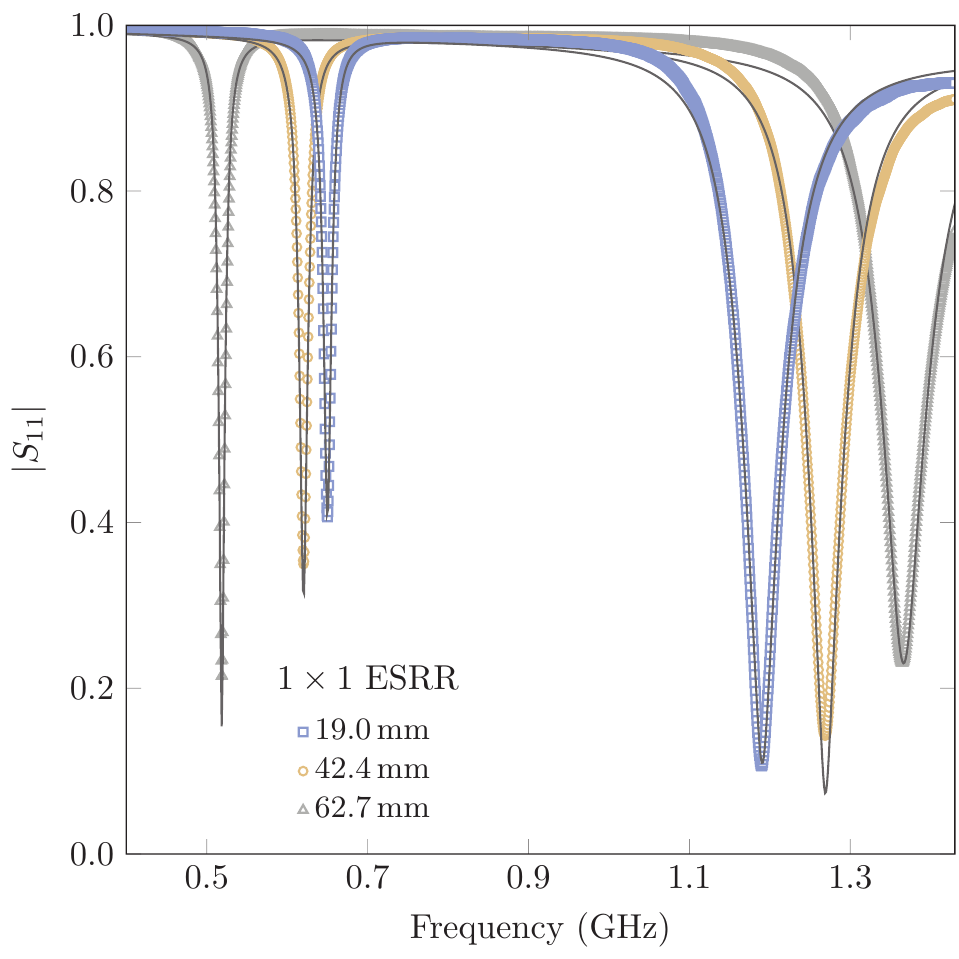} & \qquad (d)\includegraphics[height=7.5cm]{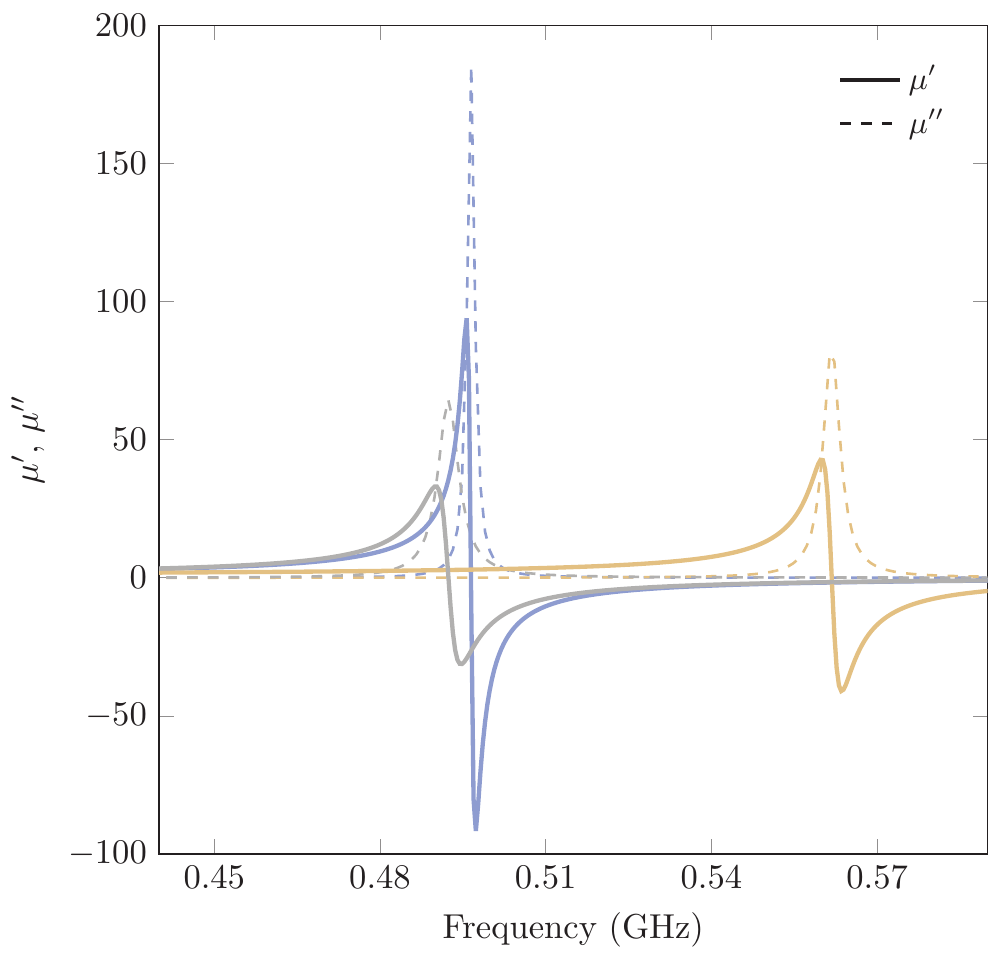}
\end{tabular}
\caption{\label{fig:Fig3}(a) $\left\vert S_{11}\right\vert$ as a function of frequency for the 1-loop-1-gap LGR when loaded with 1D arrays of $N$ SRRs.  The spacing between adjacent SRR planes was $a=\SI{21}{\milli\meter}$.  The solid lines are fits to the data used to extract the permeability of the 1D array.  The dashed line is a fit to the $N=4$ data when using fixed values of $L_1$ and $M_0$. (b) The extracted complex permeabilities of the 1D SRR arrays. (c) $\left\vert S_{11}\right\vert$ as a function of frequency for the 1-loop-1-gap LGR when loaded with single ESRRs of different lengths. (d) The extracted complex permeabilities of the ESRRs.}
\end{figure*}

\begin{figure*}
\begin{tabular}{cc}
(a)\includegraphics[height=7.5cm]{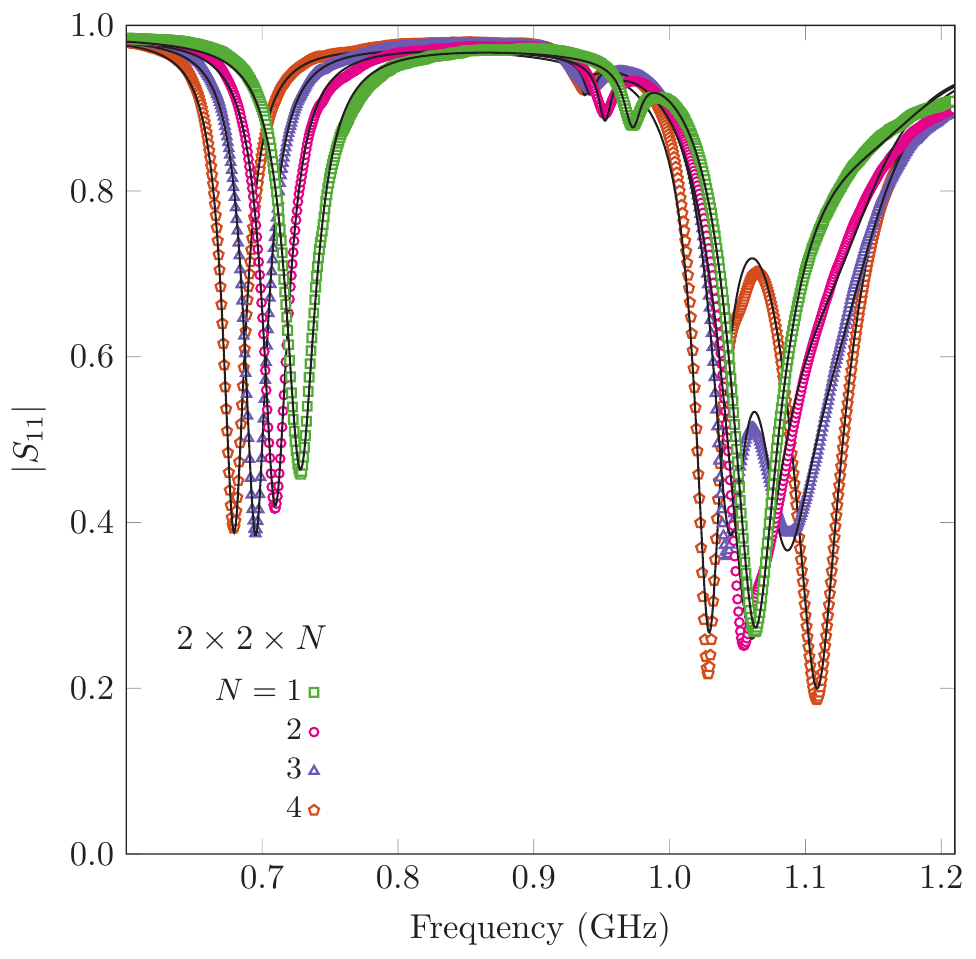} & \qquad (b)\includegraphics[height=7.5cm]{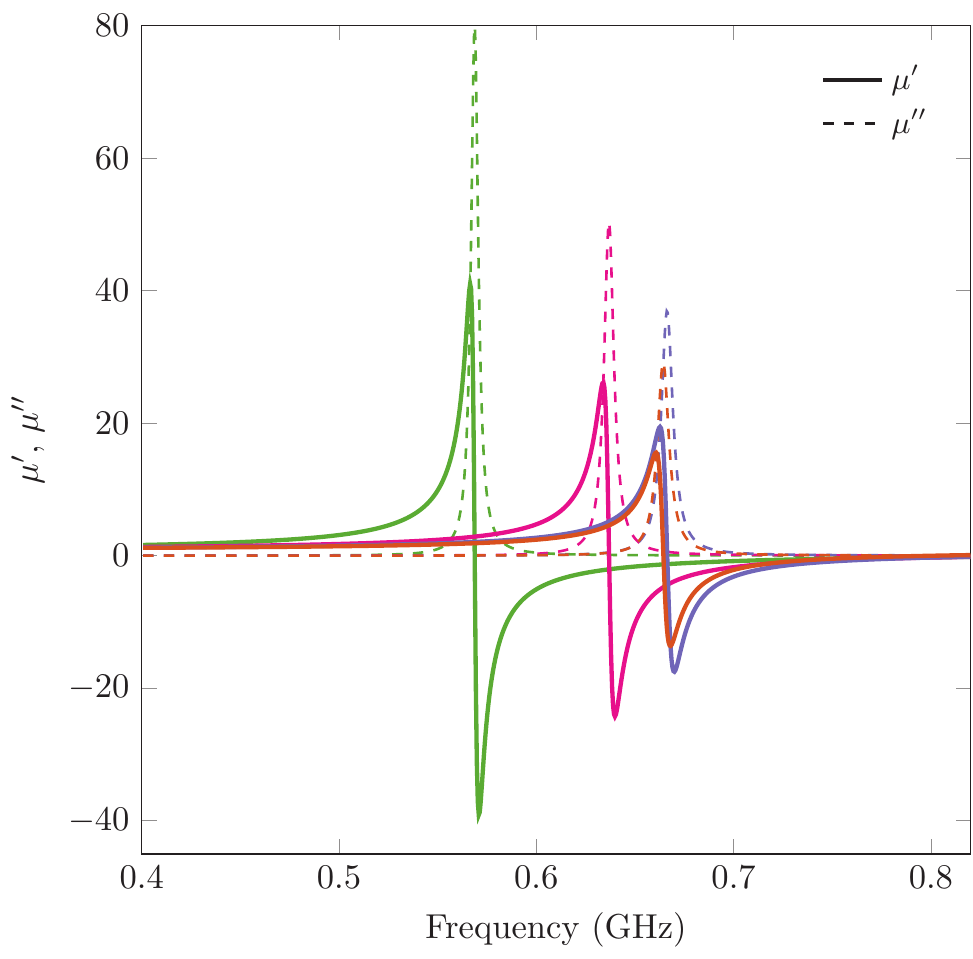}\\
(c)\includegraphics[height=7.5cm]{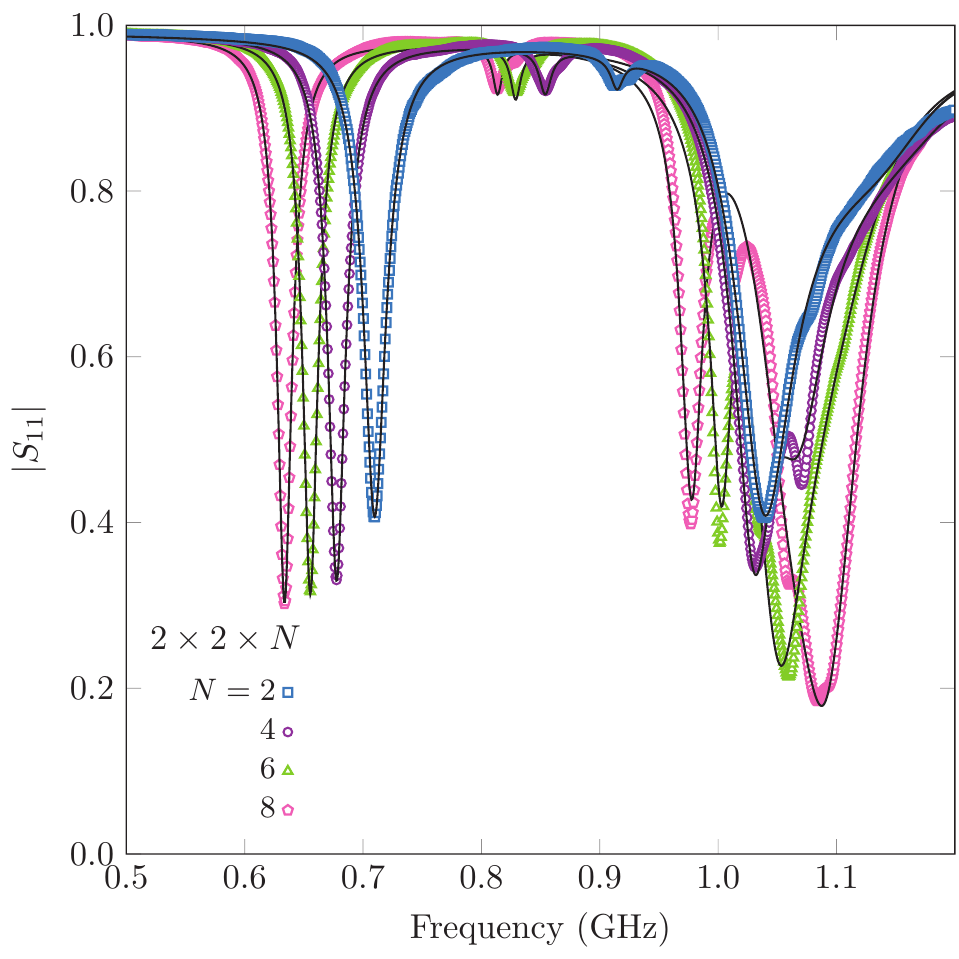} & \qquad (d)\includegraphics[height=7.5cm]{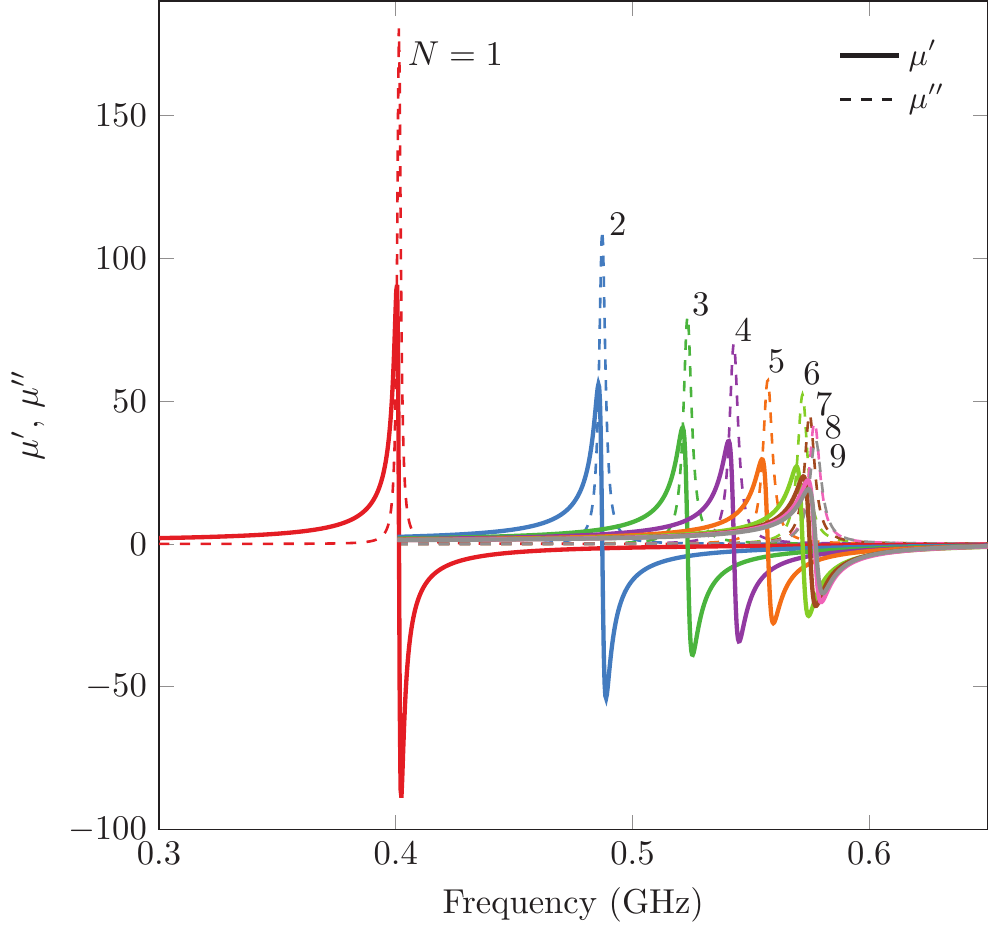}\\
(e)\includegraphics[height=7.5cm]{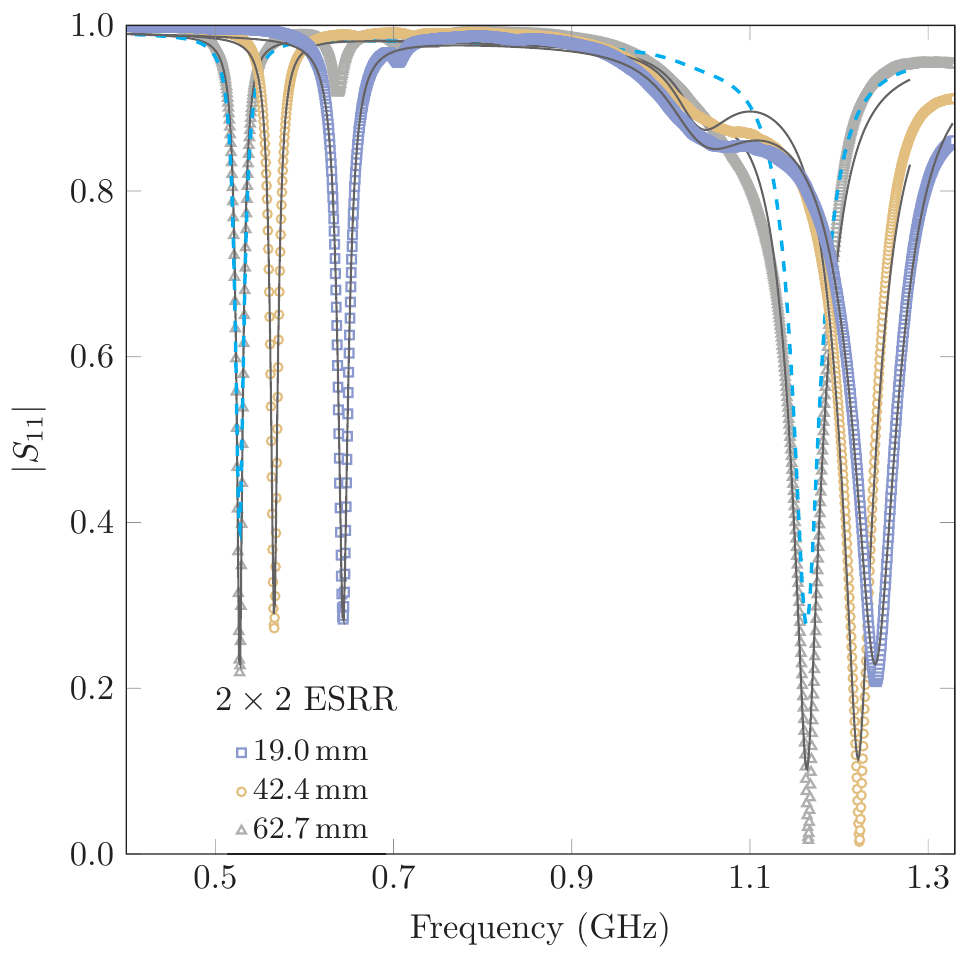} & \qquad (f)\includegraphics[height=7.5cm]{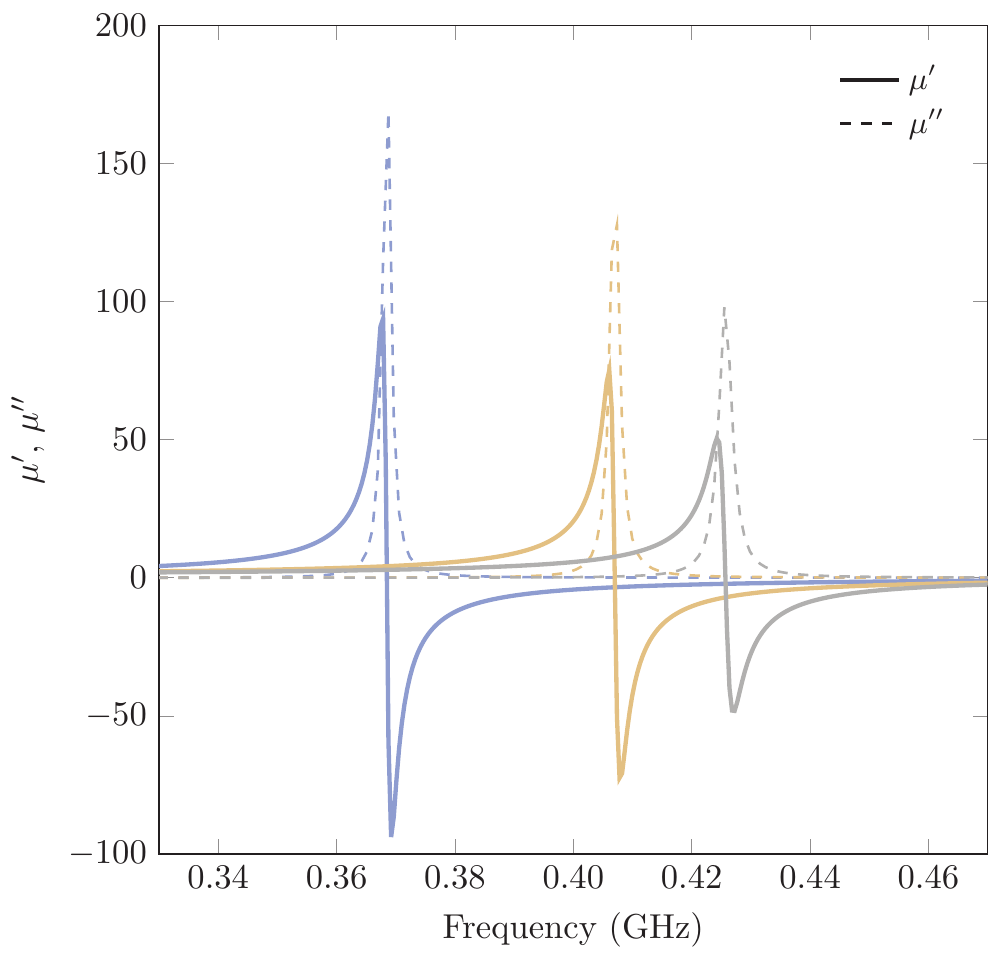}
\end{tabular}
\caption{\label{fig:Fig4}(a) $\left\vert S_{11}\right\vert$ as a function of frequency for the 2-loop-1-gap LGR when loaded with $2\times 2\times N$ arrays of SRRs.  The spacing between adjacent SRR planes was $a=\SI{21}{\milli\meter}$.  The lines are fits to the data.  (b) The complex permeabilities extracted from the data in (a). (c) Same as (a), but with an SRR plane spacing of \SI{7.8}{\milli\meter}.  For clarity, only the data for $N=2$, 4, 6, and 8 are shown. (d) The complex permeabilities extracted from the $2\times 2\times N$ $\left\vert S_{11}\right\vert$ data with an SRR plane spacing of \SI{7.8}{\milli\meter} for $N=1$ to 9. (e) $\left\vert S_{11}\right\vert$ as a function of frequency for the 2-loop-1-gap LGR when loaded with $2\times 2$ assemblies of ESRRs of different lengths.  The solid lines are fits used to determine the permeabilities of the ESRRs.  The dashed line is a fit to the $\ell_\mathrm{ESRR}=\SI{62.7}{\milli\meter}$ data when using fixed values of  $L_1$ and $M_0$. (f) The complex permeabilities extracted from the data in (e).}
\end{figure*}

\begin{figure*}
\begin{tabular}{cc}
(a)\includegraphics[height=7.5cm]{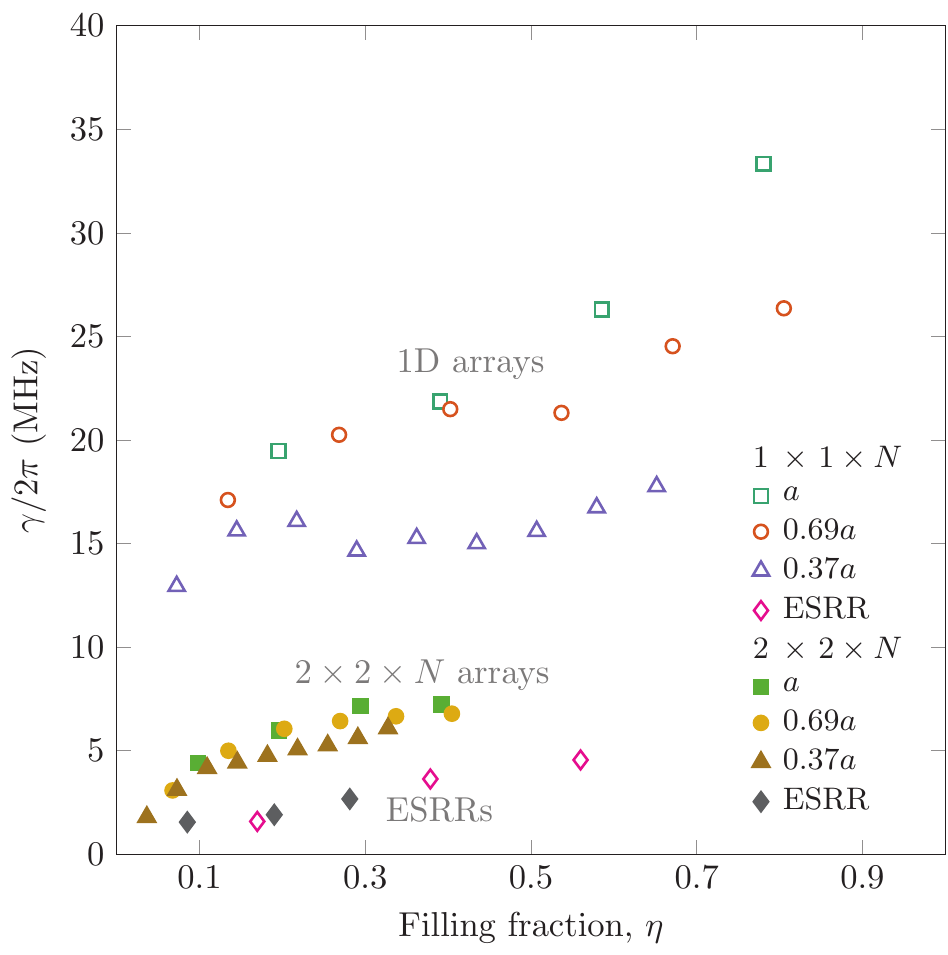} & \qquad (b)\includegraphics[height=7.5cm]{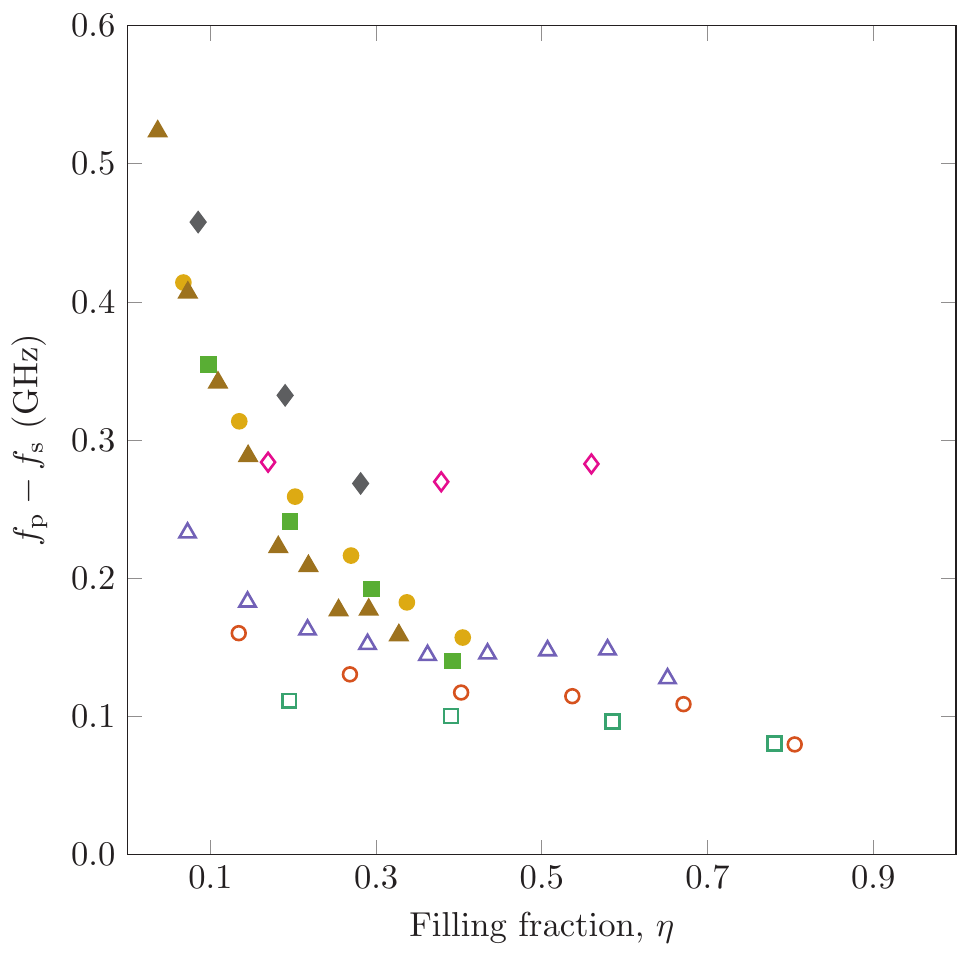}
\end{tabular}
\caption{\label{fig:Fig5}The $\chi_{\mathrm{eff},1}$ fit parameters from Table~\ref{tab:Tab2} for SRR arrays and ESRRs. (a) The damping constant $\gamma/\left(2\pi\right)$ as a function of filling fraction $\eta$. (b) The difference $f_\mathrm{p}-f_\mathrm{s}$, which determines the frequency range over which $\mu^\prime$ is negative, plotted as a function of filling fraction $\eta$.  In both (a) and (b), open symbols correspond to measurements made using the 1-loop-1-gap LGR and data obtained using the 2-loop-1-gap LGR are shown using solid symbols.  ESRR results are shown using diamonds.  In the legend, $a$, $0.69a$, and $0.37a$ refers to the spacing between adjacent planes of SRRs.}
\end{figure*}

In principle, with $L_1$, $M_0^2/R_0$, $f_0$, and $Q_0$ known from previous measurements, the parameters that determine $\left\vert S_{11}\right\vert$ of the (E)SRR-loaded LGR are the filling factor $\eta$ and the parameters that characterize the permeability of the metamaterial which, as shown in (\ref{eq:1}), are $f_\mathrm{s}$, $f_\mathrm{p}$, and $\gamma$. In our fits to the $\left\vert S_{11}\right\vert$ data of the loaded LGR, the value of $\eta$, determined by the physical size of the SRR array/ESRR, was fixed.  For the 1-loop-1-gap LGR, the filling factor was simply determined using \mbox{$\eta=\ell_\mathrm{(E)SRR}/\ell_\mathrm{LGR}$}, where the length of an array of $N$ SRRs is given by \mbox{$\ell_\mathrm{SRR}=N\left(t_\mathrm{SRR}+\ell_\mathrm{sp}\right)$}.  Here, $t_\mathrm{SRR}=\SI{1.54}{\milli\meter}$ is the thickness of the printed circuit board and $\ell_\mathrm{sp}$ is the length of the paper Bakelite spacers.  As described in Section~\ref{sec:3D} below, the 2-loop-1-gap LGR was modeled as a toroidal LGR.  For this reason, the effective bore length was assumed to be equal to twice the physical length of the resonator: $\ell_\mathrm{LGR}=\SI{224}{\milli\meter}$.

In practice, we found that fits to the $\left\vert S_{11}\right\vert$ data could be improved if we allowed $L_1$ and $M_0^2/R_0$, previously determined from empty-LGR measurements, to be adjusted.  This was accomplished by inserting scaling factors $S_{L_1}$ and $S_{M_0}$ into the loaded-LGR model.  The need to scale $L_1$ and $M_0$ is partially attributed to the fact that the SRR-loaded LGR can only be approximately modeled using an equivalent lumped-element circuit.  Furthermore, our model does not include the possibility of direct coupling between the coupling loop and the SRR array/ESRR which is expected to be important when the filling factor is large.  It is important to emphasize, however, that the conclusions of this paper are not affected by the scaling factors.  The introduction of the scaling factors, in some cases, allowed the best-fit curve to better match the depth of the $\left\vert S_{11}\right\vert$ minima at the low- and high-frequency resonances.  However, the extracted permeabilities did not depend significantly on the values of the scaling factors. 

To highlight this point, Fig.~\ref{fig:Fig3}(a) shows the $\left\vert S_{11}\right\vert$ data for a 1D array of four SRRs fit to two models.  The first model includes the $L_1$ and $M_0$ scaling factors (solid black line) and the second model fixes $S_{L_1}=1$ and $S_{M_0}=1$ (dashed cyan line).  In this case, the two best-fit lines are very similar.  The values of the scaling factors extracted from the fit were $S_{L_1}=0.87$ and $S_{M_0}=0.84$.  The extracted values of $f_\mathrm{s}$, $f_\mathrm{p}$, and $\gamma$ from the two fits differ at most by only \SI{1.3}{\percent}.  Figure~\ref{fig:Fig4}(e) shows the two models fit to the $\left\vert S_{11}\right\vert$ versus frequency data of the 2-loop-1-gap LGR when loaded with four \SI[number-unit-product=\text{-}]{63}{\milli\meter} long ESRRs.  In this case, the model with the scaling factors ($S_{L_1}=2.0$ and $S_{M_0}=2.0$) fits the data significantly better.  The additional fit parameters allow the black curve to more closely match the depth of the low- and high-frequency resonances.  We found that introducing the scaling factors caused $f_\mathrm{s}$, $f_\mathrm{p}$, and $\gamma$ to decrease by \SI{3}{\percent}, \SI{15}{\percent}, and \SI{8}{\percent}, respectively.  The observed changes in $\mu^\prime$ and $\mu^{\prime\prime}$ due to the inclusion of $S_{L_1}$ and $S_{M_0}$ are small enough that the overall trends in the permeability data are not affected.

In the analysis presented in the following sections, the scaling factors have been used to produce high-quality fits to the measured data while still generating reliable $\mu^\prime$ and $\mu^{\prime\prime}$ results.  Generally, we found that the values of $S_{L_1}$ and $S_{M_0}$ remained stable as the number of SRRs $\left(N\right)$ in an array was increased (see Table~\ref{tab:Tab2}).  The only exceptions were at large filling factors in the 1D arrays which were likely the result of direct coupling between the coupling loop and nearby SRRs.  The effect of direct coupling was not seen in the 2-loop-1-gap LGR because the SRR arrays were loaded in the bore that did not contain the coupling loop.  All of these observations are consistent with the conjecture that the scaling factors help to compensate for the limitations of the lumped-element circuit model of the LGR.

\subsection{\label{sec:1D}1D SRR Arrays}
Figure~\ref{fig:Fig3}(a) shows the measured frequency dependence of $\left\vert S_{11}\right\vert$ for the 1-loop-1-gap LGR when loaded with a 1D array of $N$ SRRs separated by distance $a$, where $1\le N\le 4$.  A major advantage of the resonant measurement technique using LGRs is that the resulting $\left\vert S_{11}\right\vert$ data are very clean.  Figure~\ref{fig:Fig3}(b) shows the frequency dependence of $\mu^\prime$ and $\mu^{\prime\prime}$ determined from the $\left\vert S_{11}\right\vert$ fit parameters. 

For all but the $N=1$ $\left\vert S_{11}\right\vert$ data, an additional dip emerges between the two dominant resonances.  This weak feature is likely caused by an additional resonant mode that emerges due to couplings between adjacent SRRs in the array.  The permeability model given in (\ref{eq:1}) will only produce an $\left\vert S_{11}\right\vert$ double resonance and is unable to capture this additional feature.  In order to properly fit the measured data, the permeability model was expanded.  In general, we assumed that \mbox{$\mu_\mathrm{eff}=1+\chi_\mathrm{eff,1}+\chi_\mathrm{eff,2}+\dots+\chi_{\mathrm{eff},n}$} which is a model that still preserves $\mu^{\prime\prime}>0$ and the Kramers-Kronig relations.  In this expression
\begin{equation}
\chi_{\mathrm{eff},n}=-\frac{1-\left(f_{\mathrm{s},n}/f_{\mathrm{p},n}\right)^2}{1-\left(f_{\mathrm{s},n}/f\right)^2-j\left[\gamma_n/\left(2\pi f\right)\right]}.\label{eq:3}
\end{equation}
In practice, we have found that $\mu^\prime$ and $\mu^{\prime\prime}$ were predominantly determined by the $n=1$ term.  As shown in Table~\ref{tab:Tab2}, all of the other terms had $f_{\mathrm{s},n}/f_{\mathrm{p},n}\lesssim 1$.  Therefore, while these terms are important for capturing features in the $\left\vert S_{11}\right\vert$ data, their contribution to $\mu^\prime-j\mu^{\prime\prime}$ is typically negligible when compared to the contribution from $\chi_{\mathrm{eff},1}$.  The only exception is the 1D $N=4$ array in Fig.~\ref{fig:Fig3}(b) which has comparable features at a pair of resonance frequencies.  For these reasons, the main conclusions of this work are based on the fit parameters obtained from the $\chi_{\mathrm{eff},1}$ term in $\mu^\prime-j\mu^{\prime\prime}$.

Figures~\ref{fig:Fig3}(c) and (d) show $\left\vert S_{11}\right\vert$ and the permeability obtained when the 1-loop-1-gap LGR was loaded with ESRRs of different lengths.  The lengths were chosen such that the filling fractions \mbox{$\eta=\ell_\mathrm{(E)SRR}/\ell_\mathrm{LGR}$} of the SRR and ESRR measurements were comparable.  In the ESRR $\left\vert S_{11}\right\vert$ data, the double resonance frequency splitting is greater and the low-frequency resonance is sharper.  These two differences correspond to a more sharply-peaked permeability and a $\mu^\prime$ that is negative over a larger frequency range.  In the limit of weak damping $\left(\gamma\ll 2\pi f_\mathrm{s}\right)$, this frequency range is given by \mbox{$f_\mathrm{p}-f_\mathrm{s}$}.

\begin{table*}
\caption{\label{tab:Tab2}Table of the parameter values extracted from fits to $\left\vert S_{11}\right\vert$ of the 1-loop-1-gap and 2-loop-1-gap LGRs when their bores were loaded with SRR arrays/ESRRs.}
\begin{tabular}{cccccccccccccccc}
 $N$ & $\eta$ & $f_{\mathrm{s},1}$ & $f_{\mathrm{p},1}$ & $\gamma_1/2\pi$ & $f_{\mathrm{s},2}$ & $f_{\mathrm{p},2}$ & $\gamma_2/2\pi$ & $f_{\mathrm{s},3}$ & $f_{\mathrm{p},3}$ & $\gamma_3/2\pi$ & $f_{\mathrm{s},4}$ & $f_{\mathrm{p},4}$ & $\gamma_4/2\pi$ & $S_{L_1}$ & $S_{M_0}$\\
~ & ~ & (GHz) & (GHz) & (MHz) & (GHz) & (GHz) & (MHz) & (GHz) & (GHz) & (MHz) & (GHz) & (GHz) & (MHz)\\
\\[-1em]
\hline\hline
\\[-1em]
\multicolumn{16}{c}{1-loop-1-gap LGR, SRR plane spacing $a$ -- Figs.~\ref{fig:Fig3}(a) and (b)}\\
\hline
\\[-1em]
 1 & 0.20 & 0.85 & 0.96 & 19 & ~ & ~ & ~ & ~ & ~ & ~ & ~ & ~ & ~ & 1.7 & 1.6 \\
 2 & 0.39 & 0.84 & 0.94 & 22 & 1.0224 & 1.0226 & 6.9 & ~ & ~ & ~ & ~ & ~ & ~ & 1.9 & 1.7\\
 3 & 0.59 & 0.83 & 0.93 & 26 & 1.009 & 1.014 & 13 & ~ & ~ & ~ & ~ & ~ & ~ & 1.6 & 1.5 \\
 4 & 0.78 & 0.85 & 0.93 & 33 & 1.015 & 1.031 & 11 & ~ & ~ & ~ & ~ & ~ & ~ & 0.87 & 0.84\\
 \hline\hline
\\[-1em]
\multicolumn{16}{c}{1-loop-1-gap LGR, SRR plane spacing $0.69a$}\\
\hline
\\[-1em]
 1 & 0.13 & 0.80 & 0.97 & 17 & ~ & ~ & ~ & ~ & ~ & ~ & ~ & ~ & ~ & 1.8 & 1.6 \\
 2 & 0.27 & 0.79 & 0.92 & 20 & 1.0272 & 1.0277 & 9.3 & ~ & ~ & ~ & ~ & ~ & ~ & 1.9 & 1.7\\
 3 & 0.40 & 0.78 & 0.90 & 21 & 1.0072 & 1.0078 & 11 & 1.025 & 1.027 & 11 & ~ & ~ & ~ & 1.9 & 1.8 \\
 4 & 0.54 & 0.78 & 0.89 & 21 & 0.988 & 0.991 & 15 & 1.028 & 1.030 & 13 & ~ & ~ & ~ & 1.6 & 1.5\\
 5 & 0.67 & 0.79 & 0.90 & 25 & 0.985 & 0.993 & 15 & 1.034 & 1.039 & 13 & ~ & ~ & ~ & 1.1 & 1.0\\
 6 & 0.81 & 0.79 & 0.87 & 26 & 0.986 & 0.994 & 16 & 1.073 & 1.092 & 24 & ~ & ~ & ~ & 0.46 & 0.25\\
  \hline\hline
\\[-1em]
\multicolumn{16}{c}{1-loop-1-gap LGR, SRR plane spacing $0.37a$}\\
\hline
\\[-1em]
 1 & 0.07 & 0.72 & 0.95 & 13 & ~ & ~ & ~ & ~ & ~ & ~ & ~ & ~ & ~ & 1.7 & 1.6 \\
 2 & 0.14 & 0.72 & 0.90 & 16 & ~ & ~ & ~ & ~ & ~ & ~ & ~ & ~ & ~ & 1.8 & 1.7\\
 3 & 0.22 & 0.71 & 0.87 & 16 & 1.01041 & 1.01043 & 4.2 & ~ & ~ & ~ & ~ & ~ & ~ & 1.9 & 1.8 \\
 4 & 0.29 & 0.70 & 0.85 & 15 & 0.9891 & 0.9899 & 11 & 1.03365 & 1.03370 & 5.5 & ~ & ~ & ~ & 1.9 & 1.8\\
 5 & 0.36 & 0.69 & 0.84 & 15 & 0.976 & 0.978 & 13 & 1.0126 & 1.0129 & 9.1 & ~ & ~ & ~ & 1.8 & 1.8 \\
 6 & 0.43 & 0.69 & 0.84 & 15 & 0.963 & 0.966 & 14 & 0.9933 & 0.9942 & 12 & ~ & ~ & ~ & 1.7 & 1.6\\
 7 & 0.51 & 0.70 & 0.84 & 16 & 0.948 & 0.953 & 14 & 0.9827 & 0.9847 & 13 & ~ & ~ & ~ & 1.4 & 1.3\\
 8 & 0.58 & 0.71 & 0.85 & 17 & 0.941 & 0.951 & 14 & 0.980 & 0.985 & 13 & ~ & ~ & ~ & 0.9 & 0.9\\
 9 & 0.65 & 0.71 & 0.84 & 18 & 0.937 & 0.948 & 14 & 0.990 & 1.005 & 28 & ~ & ~ & ~ & 0.55 & 0.56\\
 \hline\hline
\\[-1em]
 \multicolumn{16}{c}{1-loop-1-gap LGR, ESSR -- Figs.~\ref{fig:Fig3}(c) and (d)}\\
\hline
\\[-1em]
~ & 0.17 & 0.50 & 0.78 & 1.6 & ~ & ~ & ~ & ~ & ~ & ~ & ~ & ~ & ~ & 1.4 & 1.5\\
~ & 0.38 & 0.56 & 0.83 & 3.6 & ~ & ~ & ~ & ~ & ~ & ~ & ~ & ~ & ~ & 1.1 & 1.2\\
~ & 0.56 & 0.49 & 0.78 & 4.6 & ~ & ~ & ~ & ~ & ~ & ~ & ~ & ~ & ~ & 0.83 & 0.88\\
 \hline\hline
\\[-1em]
 \multicolumn{16}{c}{2-loop-1-gap LGR, SRR plane spacing $a$ -- Figs.~\ref{fig:Fig4}(a) and (b)}\\
\hline
\\[-1em]
 1 & 0.10 & 0.57 & 0.92 & 4.4 & 0.9742 & 0.9744 & 15 & 1.142 & 1.144 & 137 & ~ & ~ & ~ & 1.4 & 1.5 \\
 2 & 0.20 & 0.64 & 0.88 & 6.0 & 0.9534 & 0.9537 & 11 & 1.107 & 1.120 & 125 & ~ & ~ & ~ & 1.4 & 1.6\\
 3 & 0.29 & 0.67 & 0.86 & 7.2 & 0.9422 & 0.9424 & 11 & 1.058 & 1.059 & 28 & 1.114 & 1.117 & 118 & 1.6 & 1.5 \\
 4 & 0.39 & 0.66 & 0.80 & 7.2 & 0.9378 & 0.9379 & 7.1 & 1.043 & 1.046 & 23 & ~ & ~ & ~ & 1.3 & 1.6\\
  \hline\hline
\\[-1em]
\multicolumn{16}{c}{2-loop-1-gap LGR, SRR plane spacing $0.69a$}\\
\hline
\\[-1em]
 1 & 0.07 & 0.51 & 0.92 & 3.1 & 0.9649 & 0.9650 & 12 & 1.132 & 1.135 & 117 & ~ & ~ & ~ & 1.3 & 1.5 \\
 2 & 0.13 & 0.58 & 0.90 & 5.0 & 0.9357 & 0.9359 & 12 & 1.115 & 1.120 & 171 & ~ & ~ & ~ & 1.5 & 1.6\\
 3 & 0.20 & 0.61 & 0.87 & 6.1 & 0.9226 & 0.9230 & 12 & 1.096 & 1.100 & 156 & ~ & ~ & ~ & 1.6 & 1.7 \\
 4 & 0.27 & 0.63 & 0.85 & 6.4 & 0.9128 & 0.913 & 10 & 1.049 & 1.050 & 24 & 1.102 & 1.104 & 123 & 1.5 & 1.7\\
 5 & 0.34 & 0.64 & 0.82 & 6.7 & 0.9058 & 0.9062 & 8.2 & 1.044 & 1.046 & 22 & 1.0966 & 1.0986 & 111 & 1.5 & 1.6\\
 6 & 0.40 & 0.64 & 0.80 & 6.8 & 0.9004 & 0.9008 & 8.1 & 1.034 & 1.040 & 22 & 1.0918 & 1.0934 & 108 & 1.4 & 1.6\\
  \hline\hline
\\[-1em]
\multicolumn{16}{c}{2-loop-1-gap LGR, SRR plane spacing $0.37a$ -- Figs.~\ref{fig:Fig4}(c) and (d)}\\
\hline
\\[-1em]
 1 & 0.04 & 0.40 & 0.92 & 1.8 & 0.96314 & 0.96317 & 13 & 1.133 & 1.134 & 112 & ~ & ~ & ~ & 1.4 & 1.5 \\
 2 & 0.07 & 0.49 & 0.89 & 3.1 & 0.9151 & 0.9153 & 13 & 1.107 & 1.110 & 150 & ~ & ~ & ~ & 1.4 & 1.6\\
 3 & 0.11 & 0.52 & 0.86 & 4.2 & 0.8740 & 0.8744 & 13 & 1.067 & 1.070 & 96 & 1.173 & 1.175 & 103 & 1.6 & 1.6 \\
 4 & 0.15 & 0.54 & 0.83 & 4.4 & 0.8543 & 0.8545 & 12 & 1.049 & 1.050 & 40 & 1.134 & 1.139 & 176 & 1.5 & 1.7\\
 5 & 0.18 & 0.56 & 0.78 & 4.8 & 0.8389 & 0.8392 & 10 & 1.0239 & 1.0246 & 23 & 1.099 & 1.102 & 105 & 1.4 & 1.7\\
 6 & 0.22 & 0.57 & 0.78 & 5.1 & 0.8292 & 0.8298 & 10 & 1.0100 & 1.0107 & 18 & 1.084 & 1.088 & 159 & 1.5 & 1.7\\
 7 & 0.25 & 0.57 & 0.75 & 5.3 & 0.8200 & 0.8205 & 9.9 & 0.9976 & 0.9991 & 18 & 1.068 & 1.070 & 86 & 1.3 & 1.6\\
 8 & 0.29 & 0.58 & 0.75 & 5.6 & 0.8140 & 0.8145 & 9.2 & 0.9887 & 0.9913 & 21 & 1.069 & 1.071 & 108 & 1.4 & 1.6\\
 9 & 0.33 & 0.58 & 0.74 & 6.1 & 0.8089 & 0.8095 & 8.7 & 0.976 & 0.980 & 21 & 1.084 & 1.085 & 71 & 1.3 & 1.6\\
  \hline\hline
\\[-1em]
 \multicolumn{16}{c}{2-loop-1-gap LGR, ESRR -- Figs.~\ref{fig:Fig4}(e) and (f)}\\
\hline
\\[-1em]
 ~ & 0.09 & 0.37 & 0.83 & 1.6 & 1.056 & 1.059 & 128 & ~ & ~ & ~ & ~ & ~ & ~ & 2.1 & 2.2 \\
 ~ & 0.19 & 0.41 & 0.74 & 1.9 & 1.052 & 1.054 & 84 & ~ & ~ & ~ & ~ & ~ & ~ & 2.2 & 2.2\\
 ~ & 0.28 & 0.43 & 0.69 & 2.7 & 1.050 & 1.051 & 81 & ~ & ~ & ~ & ~ & ~ & ~ & 2.0 & 2.0 
\end{tabular}
\end{table*}

\subsection{\label{sec:3D}$2\times 2\times N$ SRR Arrays}
The measurements of Section~\ref{sec:1D} were repeated using a 2-loop-1-gap LGR partially loaded with a \mbox{$2\times 2\times N$} array of SRRs and sets of four ESRRs (in a $2\times 2$ arrangement) of different lengths.  The results are shown in Fig.~\ref{fig:Fig4}.  In Figs.~\ref{fig:Fig4}(a) and (b) the spacing between adjacent planes of SRRs was $a$ and in Figs.~\ref{fig:Fig4}(c) and (d) the spacing was $0.37a$.  To achieve similar filling factors, arrays with reduced spacings required a larger number of SRRs.  When working with SRR plane spacings of $a$, $0.69a$ and $0.37a$, the number of SRR planes in the largest arrays investigated were 4, 6, and 9, respectively.  Note that, for clarity, only a subset of the $\left\vert S_{11}\right\vert$ measurements are shown in Fig.~\ref{fig:Fig4}(c).  

In the 2-loop-1-gap LGR, magnetic field lines form closed loops by passing from one bore of the resonator to the other~\cite{Froncisz:1986}.  In this way, there is very little magnetic flux outside of the resonator and it behaves very much like a toroidal LGR and, therefore, can also be modeled as an inductively-coupled $LRC$ circuit~\cite{Bobowski:2016}.  Interestingly, comparing Figs.~\ref{fig:Fig3}(b) and \ref{fig:Fig4}(b) shows the transition from a 1D array of $N$ SRRs to a \mbox{$2\times 2\times N$} array results in a much more sharply-peaked permeability.  Moreover, both sets of data exhibit more extreme values of $\mu^\prime$ (positive and negative) than have been reported in previous experimental measurements~\cite{Alitalo:2013, Itoh:2018, Liu:2016, Zhang:2008, Chen:2006}.  The suppressed damping constant is likely due to the fact that the SRRs are effectively enclosed by a conducting shield that inhibits radiative losses that would be expected in an open, or unshielded, system~\cite{Bobowski:2013, Bobowski:2016}.  The results from the $2\times 2$ ESRR structures are shown in Figs.~\ref{fig:Fig4}(e) and (f).  Remarkably, the $2\times 2$ SRR and ESRR data are much more similar to one another than was found in the case of the 1D SRRs and single ESRRs.

\section{\label{sec:discussion}Discussion}
Although not shown, we have also measured 1D arrays of SRRs in which the spacings between the SRR planes were $0.69a$ and $0.37a$ and $2\times 2\times N$ arrays in which the spacing was $0.69a$.  The data for the 1D arrays follow the same trends shown in Figs.~\ref{fig:Fig3}(a) and (b) for the arrays with an SRR spacing of $a$.  Likewise, the $2\times 2\times N$ arrays with a spacing of $0.69a$ are qualitatively similar to the data presented in Figs.~\ref{fig:Fig4}(a) -- (d).  We note that the measured $\left\vert S_{11}\right\vert$ data exhibit the features predicted in the theoretical considerations presented in~\cite{Bobowski:2018}.  In particular, we observe two main resonances with the first being somewhat near the LGR resonant frequency $f_0$ and the second at a higher frequency.  The parameters extracted from fits to all of the $\left\vert S_{11}\right\vert$ measurements are summarized in Table~\ref{tab:Tab2}.  Figures~\ref{fig:Fig5}(a) and (b) show the permeability parameters extracted from all SRR array and ESRR $\left\vert S_{11}\right\vert$ fits as a function of the LGR bore filling fraction.  In these plots, we show only the parameters extracted from the dominant $\chi_\mathrm{eff,1}$ contribution to \mbox{$\mu^\prime-j\mu^{\prime\prime}$}.  The uncertainties in the best-fit parameters are smaller than the data points used in the figures.  The scatter in the data is predominantly due to small variations in the parallelism and spacing of the SRRs in the arrays.

First, the damping constant results are considered.  In the original calculation of Pendry {\it et al}., the damping constant of an ESRR was shown to be $\gamma=\omega \delta/r$ where \mbox{$\omega=2\pi f$}, \mbox{$r=\left(r_1+r_2\right)/2$} is the average radius of the ESRR, \mbox{$\delta=\sqrt{2\rho/\left(\mu_0\omega\right)}$} is the EM skin depth, and $\rho$ is the resistivity of copper~\cite{Pendry:1999, Bobowski:2018}. Taking $\rho=\SI{1.7e-8}{\ohm\meter}$ and $f=\SI{0.5}{\giga\hertz}$, Pendry's theory predicts $\gamma/\left(2\pi\right)\approx\SI{0.16}{\mega\hertz}$ for the ESRRs used in our experiments.  As shown in Table~\ref{tab:Tab2}, for both the 1-loop-1-gap and 2-loop-1-gap LGRs, the shortest ESRRs ($\ell_\mathrm{ESRR}=\SI{19}{\milli\meter}$) resulted in $\gamma=\SI{1.6}{\mega\hertz}$ which is an order magnitude greater than the expected value.

One reason for the difference is that the calculated value does not take into account the loss tangent of the paper Bakelite tube between the concentric copper sheets.  Furthermore, the calculated damping constant does not consider EM radiative losses~\cite{Bobowski:2013, Bobowski:2016}.  Placing the ESRR inside the bore of a conducting LGR is expected to suppress the radiative losses.  As the length of the ESRR is increased, one expects the shielding by the LGR bore to become less effective and, hence, the damping constant to increase.  The diamond data points in Fig.~\ref{fig:Fig5}(a) show that $\gamma/\left(2\pi\right)$ increases approximately linearly with filling fraction. Furthermore, the $\gamma$ values extracted from measurements made using the 1-loop-1-gap and 2-loop-1-gap LGRs follow a consistent trend.

Pendry's calculation for the SRR array damping constant showed that $\gamma\propto\ell_\mathrm{SRR}$~\cite{Pendry:1999}.  Therefore, one expects $\gamma$ to increase as the spacing between adjacent SRR planes is increased.  Our analysis of 1D and $2\times 2\times N$ SRRs shows that, for similar values of $\eta$, $\gamma$ increases as the SRR spacing is increased from $0.37a$ to $a$ (see Table~\ref{tab:Tab2}).  However, it is immediately clear that the 1D SRR arrays have values of $\gamma$ that are up to an order of magnitude greater than those of the ESRRs.  On its own, this observation is not remarkable.  After all, the SRRs were fabricated on an FR-4 dielectric while the ESRRs were made using paper Bakelite tubes.  These materials will have different loss tangents and there is no reason to expect the two structures to have similar $\gamma$ values.

The most striking observation is that the $2\times 2\times N$ SRR arrays have a damping constant that is significantly lower than that of the 1D arrays.  We speculate that the reason for the dramatic difference is that, in the case of the $2\times 2\times N$ array, the magnetic flux from a single SRR can couple to resonators in adjacent planes as well as to neighboring resonators within its own plane.  In contrast, an SRR in a 1D array can only couple to resonators in adjacent planes.  As a result, the 1D array is less efficient at exchanging energy between resonant elements within the metamaterial structure and hence exhibit enhanced losses.  If this argument is correct, then $\gamma$ would be expected to be even lower in $3\times 3\times N$ arrays of SRRs.  The importance of coupling between SRRs has been noted and systematically studied by Gay-Balmaz and Martin in structures that contained up to six resonators~\cite{Gay-Balmaz:2002}.

Recall that $f_\mathrm{p}-f_\mathrm{s}$ determines the range of frequencies for which $\mu^\prime<0$.  Figure~\ref{fig:Fig5}(b) shows that there is reasonably good consistency between the $f_\mathrm{p}-f_\mathrm{s}$ results obtained from the 1D and \mbox{$2\times 2\times N$} SRR arrays.  These values, however, fall below what was found for the ESRRs (diamond data points). For sufficiently large arrays $\left(\eta\gtrsim 0.4\right)$, we find that $f_\mathrm{p}-f_\mathrm{s}$ is approximately constant, which is one of the predictions by Pendry {\it et al}.  Another of their predictions is that $f_\mathrm{s}\propto\sqrt{\ell_\mathrm{SRR}}$~\cite{Pendry:1999}.  The $f_{\mathrm{s},1}$ column of Table~\ref{tab:Tab2} clearly shows an SRR resonant frequency that decreases as the spacing between the SRR planes is reduced.  This observation holds for both the 1D and $2\times 2\times N$ arrays.

For the $2\times 2\times N$ arrays, Figs.~\ref{fig:Fig4}(b) and (d) for array spacings of $a$ and $0.37a$ show that $\mu^\prime$ and $\mu^{\prime\prime}$ evolve in a systematic way as the number of SRR planes $\left(N\right)$ in the array is increased.  In particular, $f_\mathrm{s}$ and $\gamma$ both increase with increasing $N$.  However, once a threshold value of $N$ is reached, the effective permeability of the array becomes relatively insensitive to further increases in the number of SRR planes.  Although the data are not shown graphically, Table~\ref{tab:Tab2} shows that $2\times 2\times N$ arrays with a spacing of $0.69a$ follow a similar trend.  For the array with an SRR spacing of $0.37a$, Fig.~\ref{fig:Fig4}(d) shows that the threshold value of $N$ is approximately six.  For the array with an SRR spacing of $a$, the threshold value of $N$ is as low as three or four.  These data suggest that, for the $2\times 2\times N$ arrays that we studied, we have reached array sizes that approximate reasonably well the behavior that would be expected for an infinitely long SRR array.

\section{\label{sec:conclusions}Conclusions}
We have developed a new experimental method for determining the complex effective permeability of microwave metamaterials.  In our method, the bore of an LGR is partially filled with the metamaterial and a VNA is used to measure the reflection coefficient from the inductively-coupled LGR.  Using an assumed model for $\mu^\prime-j\mu^{\prime\prime}$ of the metamaterial, the $\left\vert S_{11}\right\vert$ versus frequency data are fit to extract the parameters characterizing the permeability.  We applied our method to 1D and $2\times 2\times N$ arrays of SRRs and compared the results to those obtained from ESRRs and to theoretical predictions.

In order to capture all of the features present in the $\left\vert S_{11}\right\vert$ measurements, we had to extend the permeability model that was developed by Pendry {\it et al}.\ for SRR arrays.  The modified model still ensures that $\mu^{\prime\prime}>0$ at all frequencies and satisfies the Kramers-Kronig relations for magnetic susceptibility.  We found, however, that the additional terms in the extended model did not contribute in a significant way to the experimentally-extracted permeabilities of the SRR arrays.

The bore of the LGR acts as an EM shield such that radiative losses from the SRR arrays were suppressed and we observed very low damping and extreme values of $\mu^\prime$ and $\mu^{\prime\prime}$.  In agreement with the theory of Pendry {\it et al}., we found SRR array resonant frequencies and damping constants that increased as the spacing between SRR planes was increased.

When compared to the 1D SRR arrays, the $2\times 2\times N$ arrays exhibited far less damping.  We speculate that this effect is the result of intraplane coupling between SRRs.  Finally, for the $2\times 2\times N$ arrays, $f_\mathrm{p}-f_\mathrm{s}$ initially decreased and $\gamma$ increased as the number of SRR planes in the array was increased.  However, once a threshold value of $N$ was reached, the extracted permeability became insensitive to further increases in the array size.  This suggests that, for the largest arrays studied, we observed the intrinsic behavior that would be expected for an infinitely long array of $2\times 2$ SRR planes.

In future work, we plan to measure the permeability of arrays of conducting rods, or cut wires, which have been reported to exhibit anti-resonant behavior when exposed to EM plane waves~\cite{Koschny:2003}.  We can also study the magnetic response of negative-index metamaterials made from a lattice that combines SRRs and cut wires.  It should also be possible to investigate \mbox{$3\times 3\times N$} metamaterial structures using a 3-loop-2-gap LGR with a larger bore size~\cite{Wood:1984, Hyde:1989}.  Finally, we point out that 3D EM simulations of our experimental setup could be used to investigate the origin of the additional resonant modes that were observed in our measurements.  Furthermore, these simulations could lead to additional insights concerning the importance of intraplane coupling in the $2\times 2\times N$ SRR arrays.

\section*{Acknowledgment}
We gratefully acknowledge the support provided by Thomas Johnson and the UBC Okanagan machine shop.  We also thank the referees for their thorough review of the manuscript and the valuable feedback provided.

\ifCLASSOPTIONcaptionsoff
  \newpage
\fi



\bibliographystyle{IEEEtran}

\begin{thebibliography}{1}




\bibitem{Pendry:1999}
J. B. Pendry, A. J. Holden, D. J. Robbins, and W. J. Stewart, ``Magnetism from conductors and enhanced nonlinear phenomena,'' {\it IEEE Trans.\ on Microw.\ Theory Tech.}, vol.~47, no.~11, pp.~2075--2084, Nov.~1999.

\bibitem{Smith:2000}
D. R. Smith, W. J. Padilla, D. C. Vier, S. C. Nemat-Nasser, and S. Schultz, ``Composite medium with simultaneously negative permeability and permittivity,'' {\it Phys.\ Rev.\ Lett.}, vol.~84, no.~18, pp.~4184--4187, May~2000.

\bibitem{Shelby:2001}
R. A. Shelby, D. R. Smith, and S. Schultz, ``Experimental verification of a negative index of refraction,'' {\it Science}, vol.~292, no.~5514, pp.~77--79, Apr.~2001.

\bibitem{Padilla:2006}
W. J. Padilla, D. N. Basov, and D. R. Smith, ``Negative refractive index metamaterials,'' {\it Metamaterials Today}, vol.~9, no.~7-8, pp.~28--35, Aug.~2006.

\bibitem{Alitalo:2009}
P. Alitalo and S. Tretyakov, ``Electromagnetic cloaking with metamaterials,'' {\it Metamaterials Today}, vol.~12, no.~3, pp.~22--29, Mar.~2009.


\bibitem{Bobowski:2018}
J. S. Bobowski, ``Probing split-ring resonator permeabilities with loop-gap resonators,'' {\it Can.\ J.\ Phys.}, vol.~96, no.~8, pp.~878--886, 2018.

\bibitem{Nicolson:1970}
A. M. Nicolson and G. F. Ross, ``Measurement of the intrinsic properties of materials by time-domain techniques,'' {\it IEEE Trans.\ Instrum.\ Meas.}, vol.~19, no.~4, pp.~337--382, Nov.~1970.

\bibitem{Weir:1974}
W. B. Weir, ``Automatic measurement of complex dielectric constant and permeability at microwave frequencies,'' {\it Proc.\ IEEE}, vol.~62, no.~1, pp.~33--36, Jan.~1974.

\bibitem{Koschny:2003}
T. Koschny, P. Marko\ifmmode \check{s}\else \v{s}\fi{}, D. R. Smith, and C. M. Soukoulis, ``Resonant and antiresonant frequency dependence of the effective parameters of metamaterials,'' {\it Phys.\ Rev.\ E}, vol.~68, no.~6, pp.~065602, Dec.~2003.

\bibitem{Liu:2016}
S.-H. Liu, L.-X. Guo, and J.-C. Li, ``Left-handed metamaterials based on only modified circular electric resonators,'' {\it J.\ Mod.\ Opt.}, vol.~63, no.~21, pp.~2220--2225, May~2016.

\bibitem{Itoh:2018}
M. Itoh, S. Yamamoto, and K. Hatakeyama, ``Experimental study of the properties of metamaterials using broadside-coupled split ring resonators,'' {\it 2018 IEEE International Symposium on Electromagnetic Compatibility and 2018 IEEE Asia-Pacific Symposium on Electromagnetic Compatibility (EMC/APEMC)}, Singapore, 2018, pp.~277--282.

\bibitem{Alitalo:2013}
P.~Alitalo, A. E. Culhaoglu, C. R. Simovski, and S. A. Tretyakov, ``Experimental study of anti-resonant behavior of material parameters in periodic and aperiodic composite materials,'' {\it J.\ Appl.\ Phys.}, vol.~113, no.~22, pp.~224903, Jun.~2013.

\bibitem{Koschny:2005}
Th. Koschny, P. Marko\ifmmode \check{s}\else \v{s}\fi{}, E. N. Economou, D. R. Smith, D. C. Vier, and C. M. Soukoulis, ``Impact of inherent periodic structure on effective medium description of left-handed and related metamaterials,'' {\it Phys.\ Rev.\ B}, vol.~71, no.~24, pp.~245105, Jun.~2005.

\bibitem{Woodley:2010}
J. Woodley and M. Mojahedi, ``On the signs of the imaginary parts of the effective permittivity and permeability in metamaterials,'' {\it J.\ Opt.\ Soc.\ Am.\ B}, vol.~27, no.~5, pp.~1016--1021, May~2010.

\bibitem{Chen:2006}
H. Chen, J. Zhang, Y. Bai, Y. Luo, L. Ran, Q. Jiang, J. A. Kong, ``Experimental retrieval of the effective parameters of metamaterials based on a waveguide method,'' {\it Opt.\ Express}, vol.~14, no.~26, pp.~12944--12949, Dec.~2006.

\bibitem{Zhang:2008}
J. Zhang, H. Chen, L. Ran, Y. Luo, B.-I. Wu, and J. A. Kong, ``Experimental characterization and cell interactions of a two-dimensional isotropic left-handed metamaterial,'' {\it Appl.\ Phys.\ Lett.}, vol.~92, no.~8, pp.~084108, Feb.~2008.

\bibitem{Marques:2003}
R. Marqu\'es, F. Mesa, J. Martel, and F. Medina, ``Comparative analysis of edge- and broadside- coupled split ring resonators for metamaterial design -- theory and experiments,'' {\it IEEE Trans.\ Antennas Propagat.}, vol.~51, no.~10, pp.~2572--2581, Oct.~2003.


\bibitem{Bobowski:2013}
J. S. Bobowski, ``Using split-ring resonators to measure the electromagnetic properties of materials: An experiment for senior physics undergraduates,'' {\it Am.\ J.\ Phys.}, vol.~81, no.~12, pp.~899--906, Dec.~2013.

\bibitem{Bobowski:2017}
J. S. Bobowski and A. P. Clements, ``Permittivity and conductivity measured using a novel toroidal split-ring resonator,'' {\it IEEE Trans.\ Microw.\ Theory Tech.}, vol.~65, no.~6, pp.~2132--2138, Jun.~2017.

\bibitem{Dubreuil:2019}
J. Dubreuil and J. S. Bobowski, ``Ferromagnetic resonance in the complex permeability of an {Fe$_3$O$_4$}-based ferrofluid at radio and microwave frequencies,'' {\it J.\ Magn.\ Magn.\ Mater.}, vol.~489, pp.~165387, Nov.~2019.

\bibitem{Hardy:1981}
W. N. Hardy and L. A. Whitehead, ``Split-ring resonator for use in magnetic resonance from 200--2000 {MHz},'' {\it Rev.\ Sci.\ Instrum.}, vol.~52, no.~2, pp.~213--216, Feb.~1981.

\bibitem{Froncisz:1982}
W. Froncisz and J. S. Hyde, ``The loop-gap resonator: {A} new microwave lumped circuit {ESR} sample structure,'' {\it J.\ Magn.\ Reson.}, vol.~47, no.~3, pp.~515--521, May~1982.

\bibitem{Bobowski:2015}
J. S. Bobowski, ``Using split-ring resonators to measure complex permittivity and permeability,'' in \textit{Proc.\ Conf.\ Lab.\ Instruct.\ Beyond the First Year College}, College Park, MD, USA, 2015, pp.~20--23.

\bibitem{Sharnoff:1964}
M. Sharnoff, ``Validity conditions for the Kramers-Kronig relations,'' {\it Am.\ J.\ Phys.}, vol.~32, no.~1, pp.~40--44, Jan.~1964.

\bibitem{Szabo:2010}
Z. Szab\'o, G.-H. Park, Ravi Hedge, and E.-P. Li, ``A unique extraction of metamaterial parameters based on Kramers-Kronig Relationship,'' {\it IEEE Trans.\ Microw.\ Theory Tech.}, vol.~58, no.~10, pp.~2646--2653, Oct.~2010.

\bibitem{Bobowski:2016}
J. S. Bobowski and H. Nakahara, ``Design and characterization of a novel toroidal split-ring resonator,'' {\it Rev.\ Sci.\ Instrum.}, vol.~87, no.~2, pp.~024701, Feb.~2016.

\bibitem{Rinard:1993}
G. A. Rinard, R. W. Quine, S. S. Eaton, and G. R. Eaton, ``Microwave coupling structures for spectroscopy,'' {\it J.\ Magn.\ Reson.\ Ser.\ A}, vol.~105, no.~2, pp.~137--144, Nov.~1993.

\bibitem{Froncisz:1986}
W. Froncisz, T. Oles, and J. S. Hyde, ``Q-band loop-gap resonator,'' {\it Rev.\ Sci. Instrum.}, vol.~57, no.~6, pp.~1095--1099, Jun.~1986.

\bibitem{Djordjevic:2001}
A. R. Djordjevic, R. M. Biljie, V. D. Likar-Smiljanic, and T. K. Sarkar, ``Wideband frequency-domain characterization of FR-4 and time-domain causality,'' {\it IEEE Trans.\ Electromagn.\ Compat.}, vol.~43, no.~4, pp.~662--667, Nov.~2001.


\bibitem{Gay-Balmaz:2002}
P. Gay-Balmaz and O. J. F. Martin, ``Electromagnetic resonances in individual and coupled split-ring resonators,'' {\it J.\ Appl.\ Phys.}, vol.~92, no.~5, pp.~2929--2936, Sep.~2002.

\bibitem{Wood:1984}
R. L. Wood, W. Froncisz, and J. S. Hyde, ``The loop-gap resonator. {II}. {C}ontrolled return flux three-loop, two-gap microwave resonators for {ENDOR} and {ESR} spectroscopy,'' {\it J.\ Magn.\ Reson.}, vol.~58, no.~2, pp.~243--253, Jun.~1984.

\bibitem{Hyde:1989}
J. S. Hyde, W. Froncisz, and T. Oles, ``Multipurpose loop-gap resonator,'' {\it J.\ Magn.\ Reson.}, vol.~82, no.~2, pp.~223--230, Apr.~1989.





\end{thebibliography}
%

%

\begin{IEEEbiography}[{\includegraphics[width=1in,height=1.25in,clip,keepaspectratio]{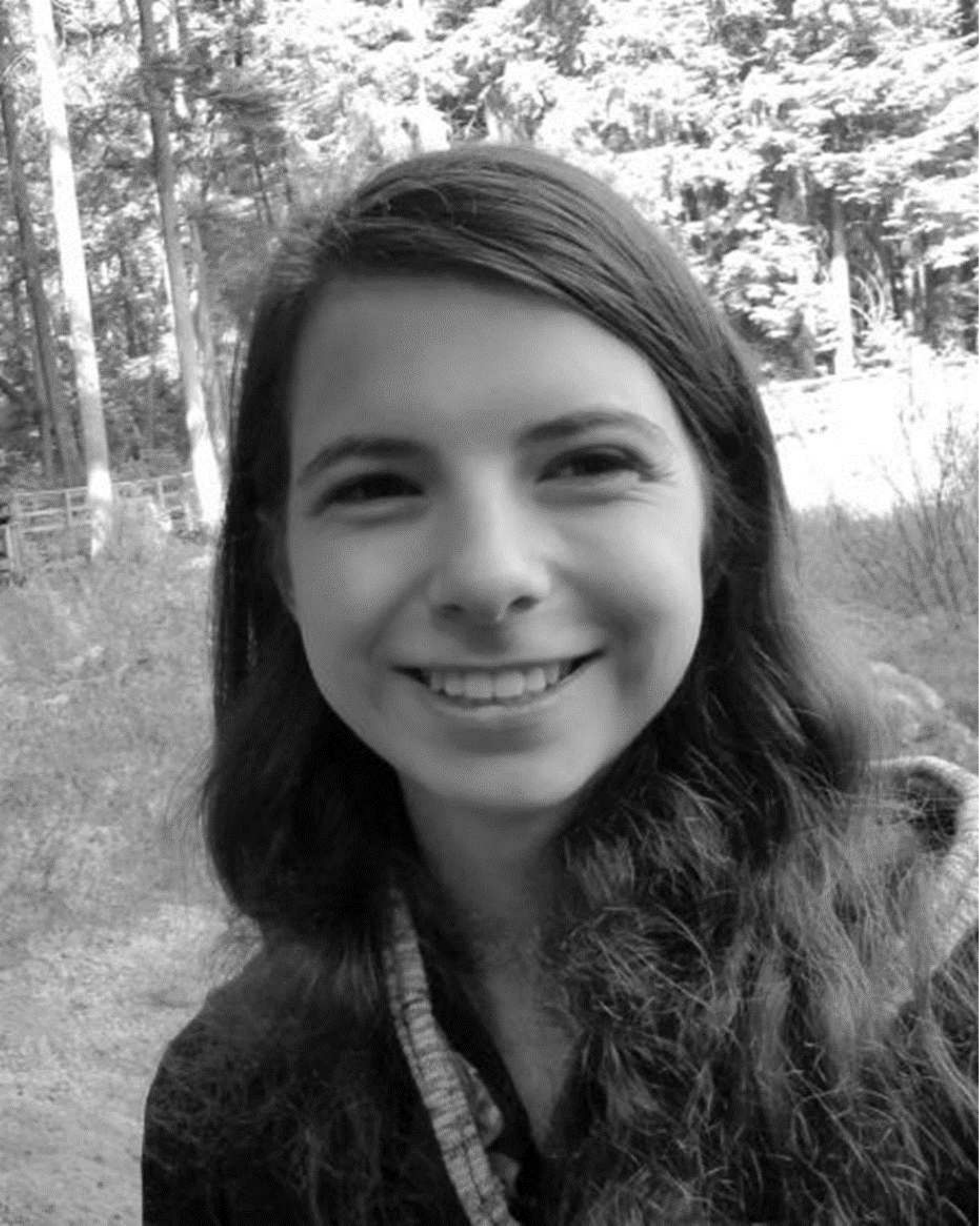}}]
{Sabrina L. Madsen}
was born in Surrey, British Columbia, Canada on August 22, 1996. She received an Honours B.Sc.\ in physics with a minor in mathematics from the Okanagan campus of the University of British Columbia, Canada in 2019. In the summer of 2018, she had a research position at the Canadian Institute of Theoretical Astrophysics at the University of Toronto, Canada studying gravitational microlensing. She is currently pursuing her Ph.D.\ in atmospheric physics at the University of Toronto. Her research interests include satellite-based spectrometry to more accurately determine the amount of carbon sequestered by vegetation and localized emissions of greenhouse gasses.
\end{IEEEbiography}

\begin{IEEEbiography}[{\includegraphics[width=1in,height=1.25in,clip,keepaspectratio]{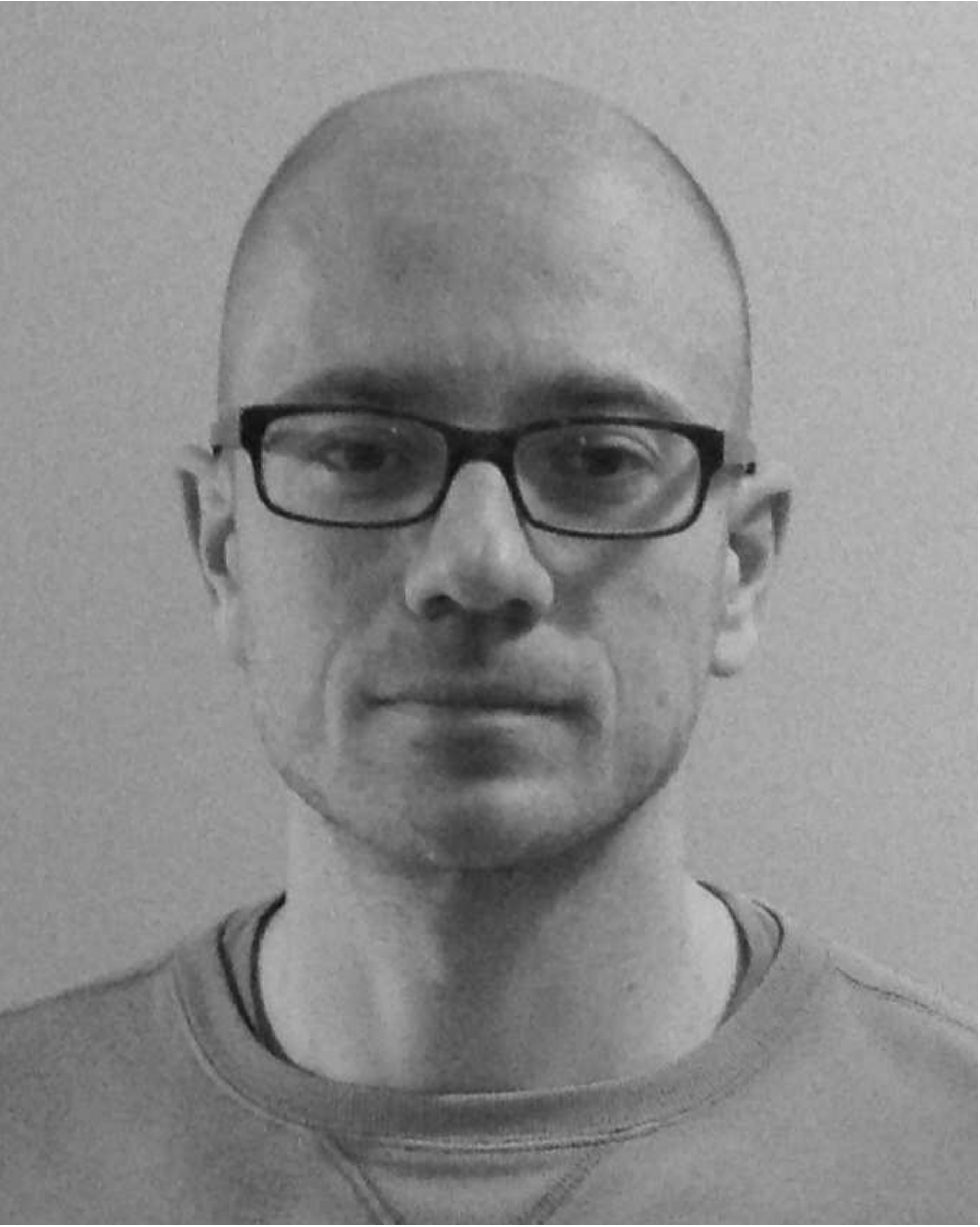}}]{Jake S. Bobowski}
was born in Winnipeg, Canada on March 9, 1979.  He received a B.Sc. degree in physics from the University of Manitoba, Canada in 2001.  He was awarded M.Sc. and Ph.D. degrees in physics from the University of British Columbia, Canada in 2004 and 2010, respectively.  From 2011 to 2012, he was a postdoctoral fellow in the RF and Microwave Technology Research Laboratory in the Department of Electrical Engineering, and is now a senior instructor in physics, at the Okanagan campus of the University of British Columbia, Canada.  He is interested in developing custom microwave techniques to characterize the electromagnetic properties of a wide range of materials.
\end{IEEEbiography}





\end{document}